\documentclass{PoS}

\title{%
\vspace*{-1cm}
\begin{minipage}{\textwidth}
\begin{flushright}
\texttt{\footnotesize
PoS(CPOD2006)002\\%
BNL-NT-07/6\\%
}
\end{flushright}
\end{minipage}\\[15pt]
QCD thermodynamics at zero and non-zero density}

\ShortTitle{QCD thermodynamics at zero and non-zero density}

\author{\speaker{Christian Schmidt}\\
        Brookhaven National Laboratory\\
        E-mail: \email{cschmidt@quark.phy.bnl.gov}}

\abstract{We present recent results on thermodynamics of QCD with almost
  physical light quark masses and a physical strange quark mass value. These
  calculations have been performed with an improved staggerd action especially
  designed for finite temperature lattice QCD. In detail we present a
  calculation of the transition temperature, using a combined chiral and
  continuum extrapolation. Furthermore we present preliminary results on the
  interaction measure and energy density at almost realistic quark
  masses. Finally we disscuss the response of the pressure to a finite quark
  chemical potential. Within the Taylor expansion formalism we calculate quark
  number susceptibilities and leading order corrections to finite chemical
  potential. This is particularly usefull for mapping out the critical region
  in the QCD phase diagram.}

\FullConference{The 3rd edition of the International Workshop --- The Critical
		 Point and Onset of Deconfinement --- \\ 
		 July 3-7 2006\\
		 Galileo Galilei Institute, Florence, Italy}

\newcommand{\nc}[1]{\newcommand{#1}}

\nc{\its}[1]{\itshape #1 \upshape}
\nc{\mc}[3]{\multicolumn{#1}{#2}{#3}}

\nc{\bc}{\begin{center}}
\nc{\ec}{\end{center}}
\nc{\ig}[1]{\bc \includegraphics{#1} \ec}

\nc{\bo}[1]{\mbox{\boldmath \( #1 \! \! \)  \unboldmath}}
\nc{\be}{\begin{eqnarray}}
\nc{\ee}{\end{eqnarray}}
\nc{\bew}{\begin{eqnarray*}}
\nc{\eew}{\end{eqnarray*}}
\nc{\bs}{\begin{subeqnarray}}   %*** requires subeqnarray.sty
\nc{\es}{\end{subeqnarray}}     %*** requires subeqnarray.sty
\nc{\nnn}{\nonumber \\}
\nc{\f}[2]{\frac{#1}{#2}}
\nc{\td}[2]{\f{d #1}{d #2}}
\nc{\pd}[2]{\f{\partial #1}{\partial #2}}
\nc{\suli}{\sum\limits}
\nc{\proli}{\prod\limits}
\nc{\ili}{\int\limits}
\nc{\sr}[2]{\stackrel{#1}{#2}}
\nc{\dps}{\displaystyle}
\nc{\ket}[1]{\left| #1 \right>}
\nc{\bra}[1]{\left< #1 \right|}
\nc{\bracket}[2]{\left< #1 \right| \left. \! #2 \right>}
\nc{\norm}[1]{\left\| #1 \right\|}
\nc{\lndm}[1]{\pd{^{#1} \ln{\det{M}}}{\mu^{#1}}}
\nc{\pdmm}[1]{M^{-1} \pd{^{#1} M}{\mu^{#1}}}
\nc{\pdm}{M^{-1}\pd{M}{\mu}}
\nc{\trac}[1]{\mbox{Tr}\left(#1\right)}
\nc{\hm}{\hat{m}}

\def\lsim{\raise0.3ex\hbox{$<$\kern-0.75em\raise-1.1ex\hbox{$\sim$}}}
\def\gsim{\raise0.3ex\hbox{$>$\kern-0.75em\raise-1.1ex\hbox{$\sim$}}}

\begin{document}

\section{Introduction}
It is by now well established that the properties of matter formed from
strongly interacting elementary particles change drastically at high
temperatures. Quarks and gluons are no longer con\-fined to move inside
hadrons but organize in a new form of strongly interacting matter,
the so-called quark-gluon plasma (QGP). The transition from hadronic
matter to the QGP as well as properties of the high temperature phase
have been studied extensively in lattice calculations over recent
years \cite{reviews}.
Nonetheless, detailed quantitative information on the transition and
the structure of the high temperature phase in the physical situation
of two light and a heavier strange quark (($2+1$)-flavor QCD)
is rare \cite{peikert_pressure,Bernard04,Bernard,aoki}.
In order to
relate experimental observables determined in relativistic heavy ion
collisions to lattice results, it is important to achieve
good quantitative control, in calculations with physical quark masses,
over basic parameters that characterize the
transition from the low to the high temperature phase of QCD. The
most fundamental quantities characterize bulk properties of hot and dense
matter: the transition temperature, energy density and pressure. 

Lattice QCD currently is the only quantitative approach to finite temperature 
QCD based on first principle calculation. At non-zero density however, lattice
QCD is harmed by the sign problem ever since its inception. To overcome the
sign problem is a challenging and outstanding problem. Nevertheless,
during the last few years a lot of progress has been made to circumvent
the sign problem for small values of $\mu_q/T$, where $\mu_q$ is the
quark chemical potential and $T$ the temperature \cite{Fodor:2004nz,
  Gavai:2003mf, methods}

\section{Lattice formulation and calculational setup}
\label{setup}

We study the thermodynamics of QCD with two light quarks
($\hm_l\equiv \hm_u= \hm_d$) and a heavier strange quark ($\hm_s$)
described by the QCD partition function which is discretized on a four
dimensional lattice of size $N_\sigma^3\times N_\tau$,
\begin{equation}
Z(\beta,\hat{m}_l,\hat{m}_s,N_\sigma,N_\tau) =
\int \prod_{x,\mu} {\rm d}U_{x,\mu}
\left( {\rm det}\; D(\hat{m}_l)\right)^{1/2}
\left( {\rm det}\; D(\hat{m}_s)\right)^{1/4} {\rm e}^{-\beta S_G (U)} \;\; .
\label{partition}
\end{equation}
Here we will use staggered fermions to discretize
the fermionic sector of QCD.
The fermions have already been integrated out, which gives
rise to the determinants of the staggered fermion matrices,
$D(\hm_l)$ and $D(\hm_s)$  for the contributions of two light and one
heavy quark degree of freedom, respectively.
Moreover, $\beta = 6/g^2$ is the gauge coupling constant, $\hat{m}_{s,l}$
denote the dimensionless, bare quark masses in units of the lattice
spacing $a$, and $S_G$ is the
gauge action which is expressed
in terms of gauge field matrices $U_{x,\mu}\in SU(3)$ located on the links
$(x,\mu)\equiv (x_0,\bf{x},\mu)$ of the four dimensional lattice;
$\mu=0,...,3$.

In our calculations we use a tree level, ${\cal O}(a^2)$ improved gauge action,
$S_G$, which includes the standard Wilson plaquette term and the
planar 6-link Wilson loop. In the fermion sector, we use an improved staggered
fermion action with 1-link and bended 3-link terms. The coefficient of the
bended 3-link
term has been fixed by demanding a rotationally invariant quark propagator
up to ${\cal O}(p^4)$, which improves the quark dispersion relation at
${\cal O}(a^2)$. This eliminates ${\cal O}(a^2)$ corrections to the pressure
at tree level and leads to a strong reduction of cut-off effects
in other bulk thermodynamic observables in the infinite temperature limit,
as well as in ${\cal O}(g^2)$ perturbation theory \cite{Heller}.
The 1-link term in the fermion action has
been `smeared' by adding a 3-link staple. This improves the flavor symmetry
of the staggered fermion action \cite{cheng}. We call this action the p4fat3
action. It has been used previously in studies of QCD thermodynamics on
lattices of temporal extent $N_\tau = 4$ with larger quark masses
\cite{peikert_pressure,peikert}. 

Our studies of the transition to the high temperature phase of QCD
\cite{RBC-Bi-Tc-2+1} have been performed on lattices of size $N_\sigma^3 \times
N_\tau$ with $N_\tau = 4$ and $6$ and spatial lattice sizes $N_\sigma = 8,~16$,
$24$ and $32$. We performed calculations for several values of the light to
strange quark mass ratio, $\hat{m}_l/\hat{m}_s$ for fixed $\hat{m}_s$. 
The strange quark mass has been chosen such that the extrapolation
to physical light quark mass values yields approximately the correct
physical kaon mass value. The range of the light quark mass corresponds to 
a regime of the pseudo-scalar (pion) mass of $150~{\rm MeV} \lsim m_{ps} \lsim
500~{\rm MeV}$   

In order to convert lattice units to physical units, zero temperature
calculations are necessary which have been performed on $16^3\times 32$ lattices.
We use parameters characterizing the shape of the static quark potential
($r_0$, $r_1$, $\sqrt{\sigma}$) as well as hadron masses to set the scale for
thermodynamic observables. 

The numerical simulation of the QCD partition function has been performed using
the RHMC algorithm \cite{rhmc}. Unlike the hybrid-R algorithm \cite{hybrid}
used in most previous studies of QCD thermodynamics performed with staggered
fermions, this algorithm has the advantage of being exact, {\it i.e.}
finite step size errors arising from the discretization of the
molecular dynamics evolution of gauge fields in configuration space are
eliminated through an additional Monte Carlo accept/reject step. This is
possible with the introduction of a rational function approximation for
roots of fermion determinants appearing in Eq.~\ref{partition}.

\section{Order parameters and susceptibilities}
To determine the QCD transition temperature and phase diagram, order parameter
of the QCD transition are indispensable.
In the chiral limit the chiral condensate $\langle
\bar{\psi}\psi \rangle$ is the order parameter for the spontaneous
chiral symmetry breaking of QCD. On the other hand in the heavy quark
limit the Polyakov loop $\langle L \rangle$ is the order parameter
of the deconfinement phase transition.
For finite quark masses, these observables remain good
indicators for the (pseudo) critical point.
Especially their susceptibilities are useful to determine the pseudo critical 
coupling $\beta_c$ in numerical simulations.

\begin{figure}[t]
\begin{center}
\begin{minipage}[c]{14.5cm}
\begin{center}
\includegraphics[width=7.0cm]{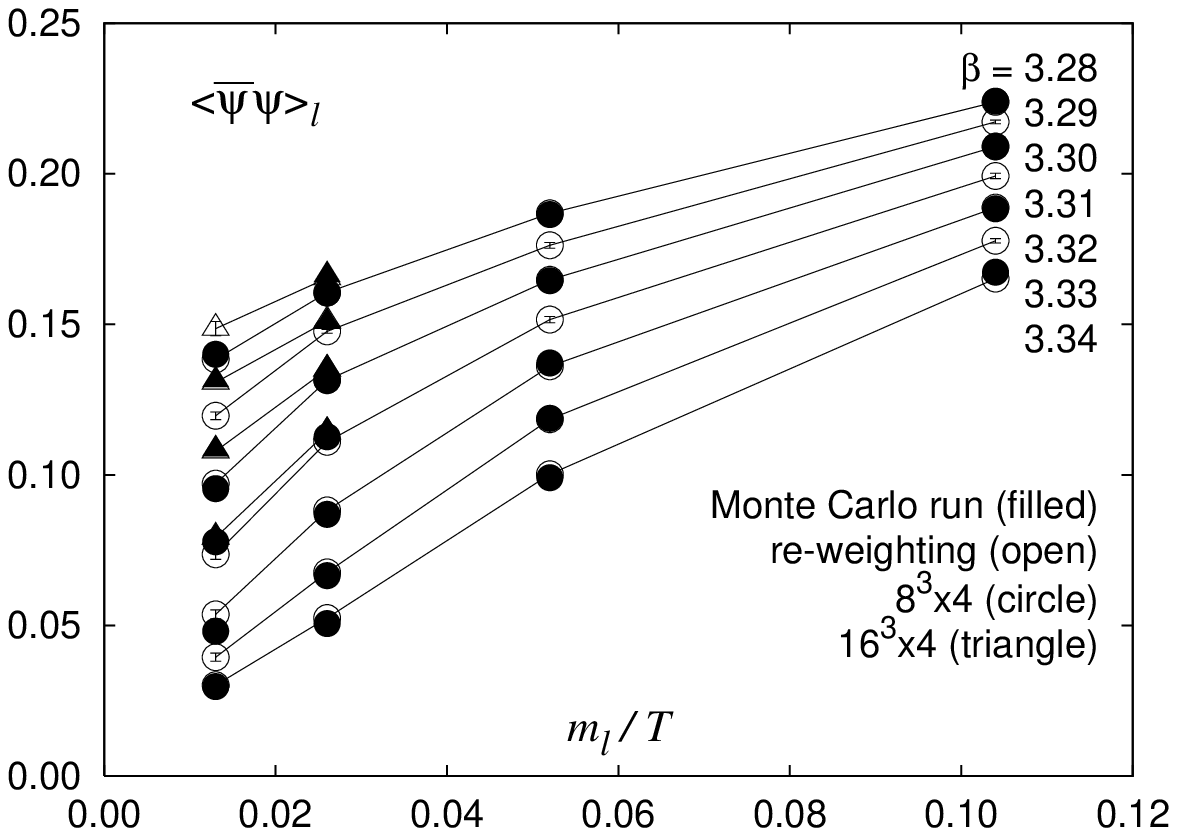}
\includegraphics[width=7.0cm]{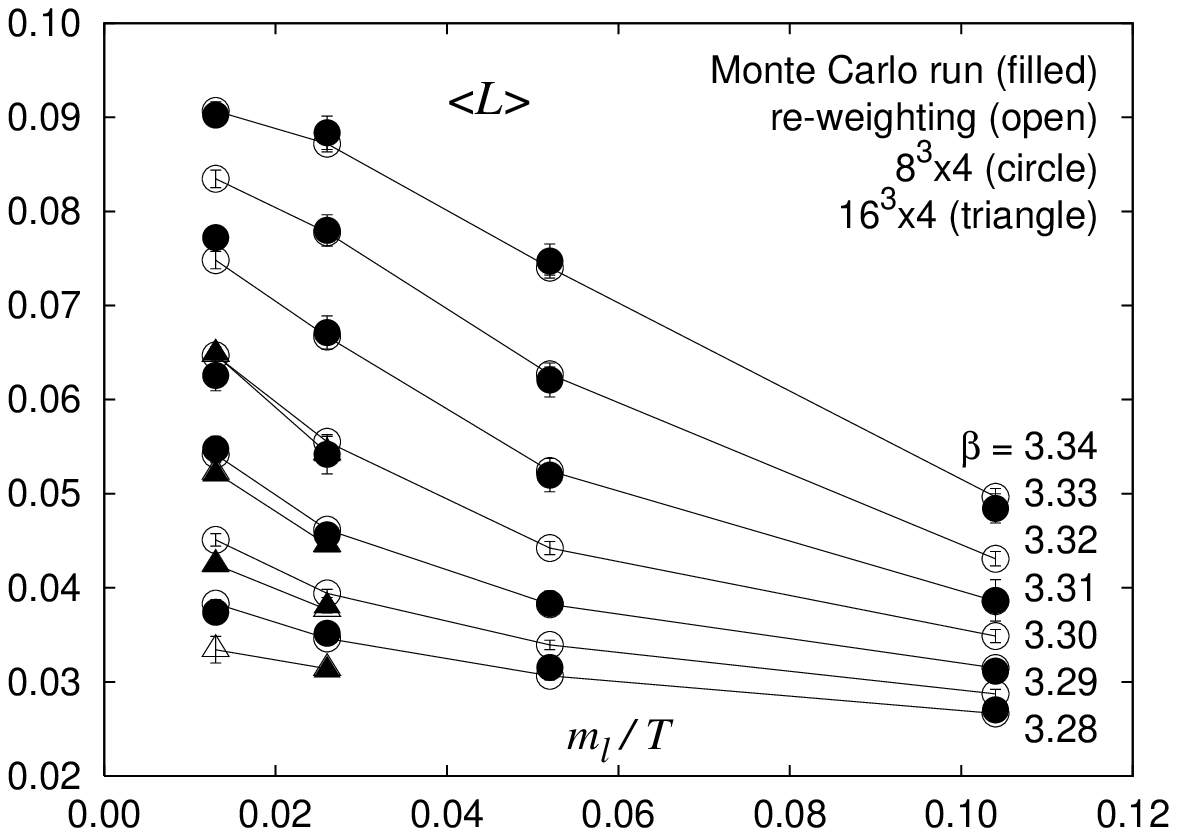}
\end{center}
\end{minipage}
\end{center}
\caption{The light quark chiral condensate in units of $a^{-3}$ (left) and
the Polyakov loop expectation value (right) as function of the bare light
quark mass in units of the temperature, $m_l/T\equiv \hm_l N_\tau$ for
fixed $\beta$ and $\hm_s=0.065$ on lattices of size
$8^3\times 4$ (circle) and $16^3\times 4$ (triangles). Shown are results
for various values of $\beta$ ranging from $\beta = 3.28$ to $\beta = 3.4$
(top to bottom for $\langle \bar{\psi}\psi\rangle$ and bottom to top for
$\langle L \rangle$). Full and open symbols show results obtained from
direct simulations and Ferrenberg-Swendsen interpolations, respectively.
}
\label{fig:pbp_4}
\end{figure}

In Figure~\ref{fig:pbp_4}(left) we compare results for the light quark
chiral condensate calculated on lattices of size $8^3\times 4$
and $16^3\times 4$. It clearly reflects the presence of finite volume
effects at small values of the quark mass. While finite volume effects seem to be
negligible for $\hm_l/\hm_s \ge 0.2$, for $\hm_l/\hm_s =0.1$ we observe
a small but statistically significant volume dependence for the chiral condensate
as well as for the Polyakov loop expectation value. This volume dependence is even
more pronounced for $\hm_l/\hm_s =0.05$ and seems to be stronger at low
temperatures.  While the value of the chiral condensate
increases with increasing volume the Polyakov loop expectation value decreases
(Figure~\ref{fig:pbp_4}(right)).

%In a theory with Goldstone bosons, e.g. in $O(N)$-symmetric spin models, it is
%expected that in the broken phase the order parameter, ${\cal O}$, changes with
%the symmetry breaking field, $h$, as ${\cal O}(h)-{\cal O}(0)\sim h^{1/2}$
%\cite{wallace}.  This behavior has also been found in QCD with adjoint quarks,
%{\it i.e.} $\langle \bar{\psi}\psi\rangle \sim c_0\; +\; c_1 (m_l/T)^{1/2}$
%\cite{lutgemeier,engels}. Our current analysis of the quark mass dependence
%of the chiral condensate is not yet accurate enough and has not yet been
%performed at small enough quark masses to verify this behavior explicitly.
%We will analyze the light quark mass limit in more detail elsewhere.

We use  the Polyakov loop susceptibility as well as the disconnected part
of the chiral susceptibility to locate the transition temperature to the
high temperature phase of QCD,
\begin{eqnarray}
\chi_L &\equiv& N_\sigma^{3}  \left(
\langle  L^2 \rangle - \langle L \rangle^2 \right) \; ,
\label{sus_L}\\
\frac{\chi_q}{T^2} &\equiv& \frac{N_\tau}{16N_\sigma^{3}} \left(
\left\langle \left( {\rm Tr}\; D^{-1}(\hm_q)\right)^2 \right\rangle -
\left\langle {\rm Tr\;} D^{-1}(\hm_q)\right\rangle^2\right) \;, \;
q\; \equiv\; l,\; s . \label{sus_chi}
\label{sus_m}
\end{eqnarray}

\begin{figure}[t]
\begin{center}
\begin{minipage}[c]{14.5cm}
\begin{center}
\includegraphics[width=7.0cm]{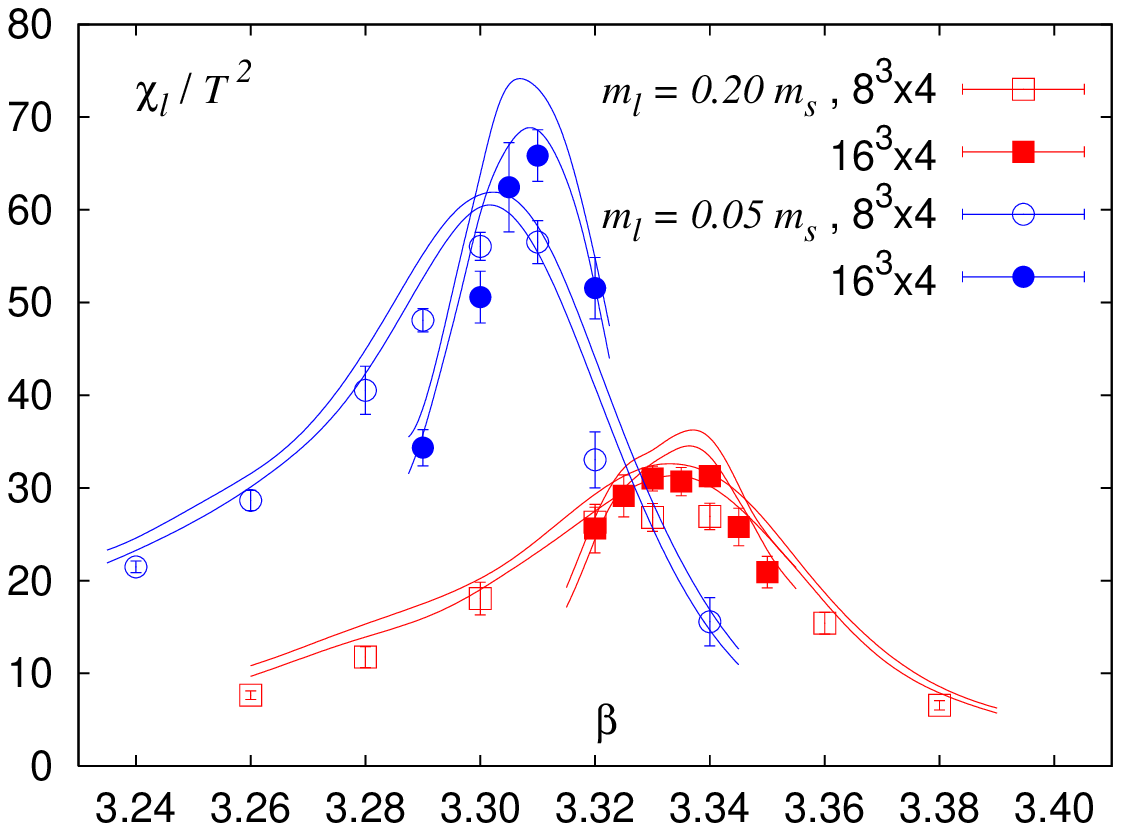}
\includegraphics[width=7.0cm]{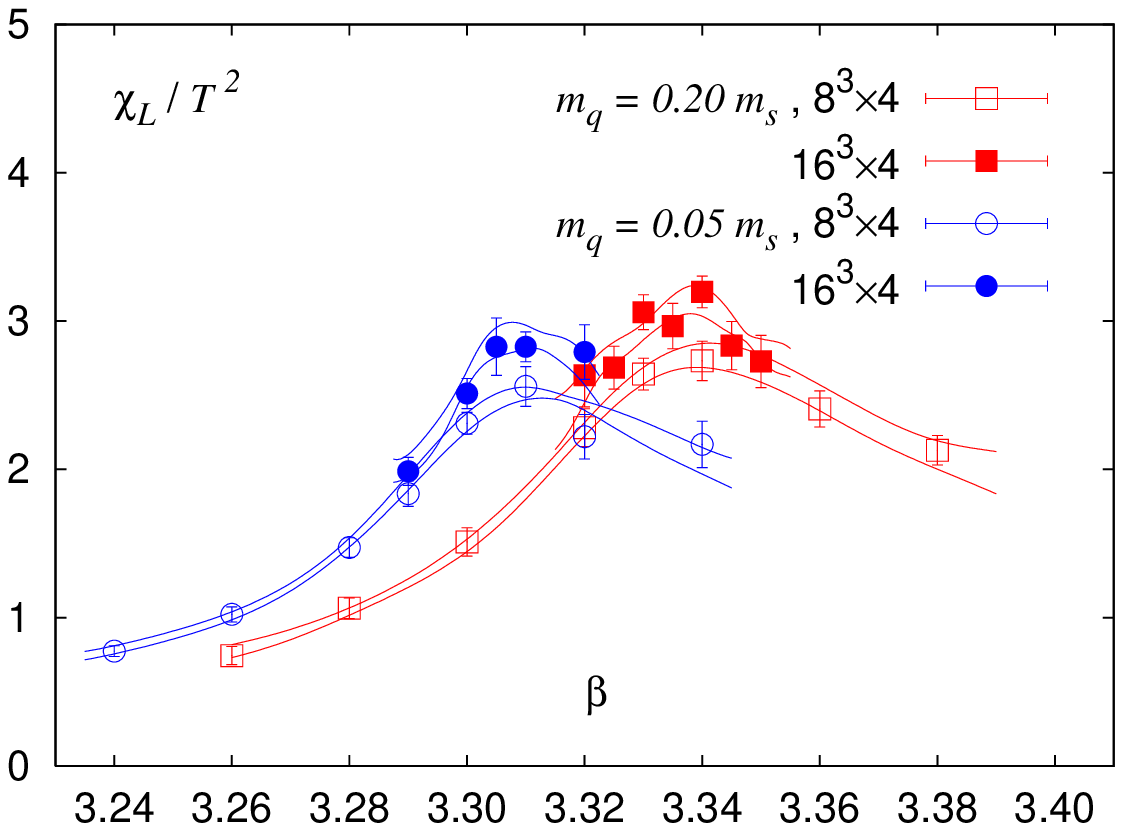}
\end{center}
\end{minipage}
\end{center}
\caption{The disconnected part of the light quark chiral susceptibility (left)
  and the Polyakov loop (right) on lattices of size $8^3\times 4$ (squares) and
  $16^3\times 4$ (circles) for two different values of the light quark
  mass. The curves show  Ferrenberg-Swendsen interpolations of the data points
  obtained from multi-parameter histograms with an error band coming from
  Ferrenberg-Swendsen reweightings performed on different jackknife samples.} 
\label{fig:chi_4}
\end{figure}

In Figure~\ref{fig:chi_4} we show results for the disconnected part of the
light quark chiral susceptibility, $\chi_l$ and the Polyakov loop, calculated
on $8^3\times 4$ and $16^3\times 4$ lattices. The location of peaks in the
susceptibilities has been determined from a Ferrenberg-Swendsen reweighting of
data in the vicinity of the peaks. Errors on the critical couplings determined
in this way have been obtained from a jackknife analysis where Ferrenberg-Swendsen
interpolations have been performed on different sub-samples.
In agreement with earlier calculations we find that the position of
peaks in $\chi_l$ and $\chi_L$ show only little volume dependence and that
the peak height changes only little, although the maxima become somewhat more
pronounced on the larger lattices. This is consistent with the transition being
a crossover rather than a true phase transition in the infinite volume limit.

Although differences in the critical coupling extracted from $\chi_L$ and
$\chi_l$ are small we find that on small lattices the peak in the Polyakov loop
susceptibility is located at a systematically larger value of the gauge
coupling $\beta$. In a finite volume this is, of course, not
unexpected, and in the infinite volume limit an ambiguity in identifying the
transition point may also remain for a crossover transition. Nonetheless, we
observe that the difference  $\beta_{c,L}-\beta_{c,l}$ decreases with
increasing
volume and is within errors consistent with zero for $16^3\times 4$, which has
the largest spatial volume expressed in units of the temperature,
$TV^{1/3}=4$. On the smallest lattice, $8^3\times 4$, we find
$\beta_{c,L}-\beta_{c,l}\simeq 0.0077(9)$. Within the statistical accuracy
of our data we also do not find any systematic quark mass dependence of this
difference, $\beta_{c,L}-\beta_{c,l}$.

%The peak positions, $\beta_c(\hm_l,\hm_s, N_\tau)$, in the chiral and Polyakov
%loop susceptibilities are generally well determined. The error on
%$\beta_{c,(L,l)}(\hm_l,\hm_s, N_\tau)$ translates, of course, into
%an uncertainty for the lattice spacing $a(\beta_c)$ which in turn contributes
%to the error on the transition temperature. In order to get a feeling for the
%accuracy required in the determination of $\beta_c$
%we give here an estimate for the dependence of the lattice cut-off
%on the gauge coupling. A shift in the gauge coupling by
%$\Delta\beta =0.02$ corresponds to a change in the lattice cut-off of
%about 5\%. An uncertainty in the determination of the critical coupling of
%about $0.01$ thus translates into a 2.5\% error on $T_c$.

In addition to the light quark condensate and its susceptibility we also
have analyzed the strange quark condensate and its susceptibility, $\chi_s$.
We find that the light and heavy quark condensates are strongly correlated,
which is easily seen in the MD-time evolution of these quantities. Already
on the smallest lattices
the position of the peak in the heavy quark susceptibilities is consistent with
that deduced from the light quark condensate. On the larger, $N_\sigma=16$,
lattices the difference  $|\beta_{c,l}-\beta_{c,s}|$ is in all cases zero
within statistical errors, which are about $3\cdot 10^{-3}$. Any
temperature difference in the crossover behavior for the light and strange
quark sector of QCD, which sometimes is discussed in phenomenological
models, thus is below the $1$~MeV level.

We also have analyzed derivatives of the QCD partition function with respect
to a quark chemical potential (see Section~\ref{sec:fluct}). We note here that
also from the analysis of the Taylor expansion coefficient $c_4$ ($d_4$) we
find values for the pseudo-critical couplings that are in agreement with the
above estimates \cite{RBC-Bi-density} (see the quartic strange quark number
fluctuations in Fig.~\ref{fig:d4} (left)). However, as mentioned above the
consistence of pseudo-critical couplings as determined from different
observables is not necessary. If the transition becomes smoother closer to the
continuum limit different observables may lead to different estimates for the
crossover point. In fact, for the stout-link improved staggered action and
lattices with temporal extent of $N_\tau=8$ to 10, a large difference in
$\beta_c$ determined from the chiral condensate and the strange quark number
fluctuations has been found \cite{fodor:tc}.

\section{Scale setting and the heavy quark potential}
In order to calculate the transition temperature in terms of an
observable that is experimentally accessible and can be used to
set the scale for $T_c$ we have to perform a zero temperature
calculation at the critical couplings $\beta_c$ determined
in the previous section. This will allow us to eliminate the unknown
lattice cut-off, $a(\beta_c)$, which determines  $T_c$  on a lattice with
temporal extent $N_\tau$, {\it i.e.} $T_c= 1/N_\tau a(\beta_c)$.
To do so we have performed calculations at zero temperature, {\it i.e.}
on lattices of size $16^3\times 32$, and calculated several hadron
masses as well as the static quark potential. From the latter we determine
the string tension and
extract short distance scale parameters $r_0,~r_1$, which are
defined as separations between the static quark anti-quark sources at which the
force between them attains certain values \cite{Sommer},
\begin{equation}
r^2\frac{{\rm d} V_{\bar{q}q}(r)}{{\rm d}r}\biggl|_{r=r_0} = 1.65   \;\; , \;\;
r^2\frac{{\rm d} V_{\bar{q}q}(r)}{{\rm d}r}\biggl|_{r=r_1} = 1.0   \; .
\label{r0}
\end{equation}
Although these scale parameters are not directly accessible to experiment they
can be well estimated from heavy quarkonium phenomenology. Moreover,
they have been determined quite accurately in lattice calculations
through  a combined analysis
of the static quark potential \cite{MILCpotential} and level splittings
in bottomonium spectra \cite{gray}. Both these calculations have been performed
on identical sets of gauge field configurations. We will use the value for
$r_0$ determined in the bottomonium calculation \cite{gray}
for all conversions of lattice results to physical units,
\begin{equation}
r_0 = 0.469(7)\;{\rm fm}  \; .
\label{r0_fm}
\end{equation}

\begin{figure}[tb]
\begin{center}
\begin{minipage}[c]{14.5cm}
\begin{center}
\includegraphics[height=5.5cm]{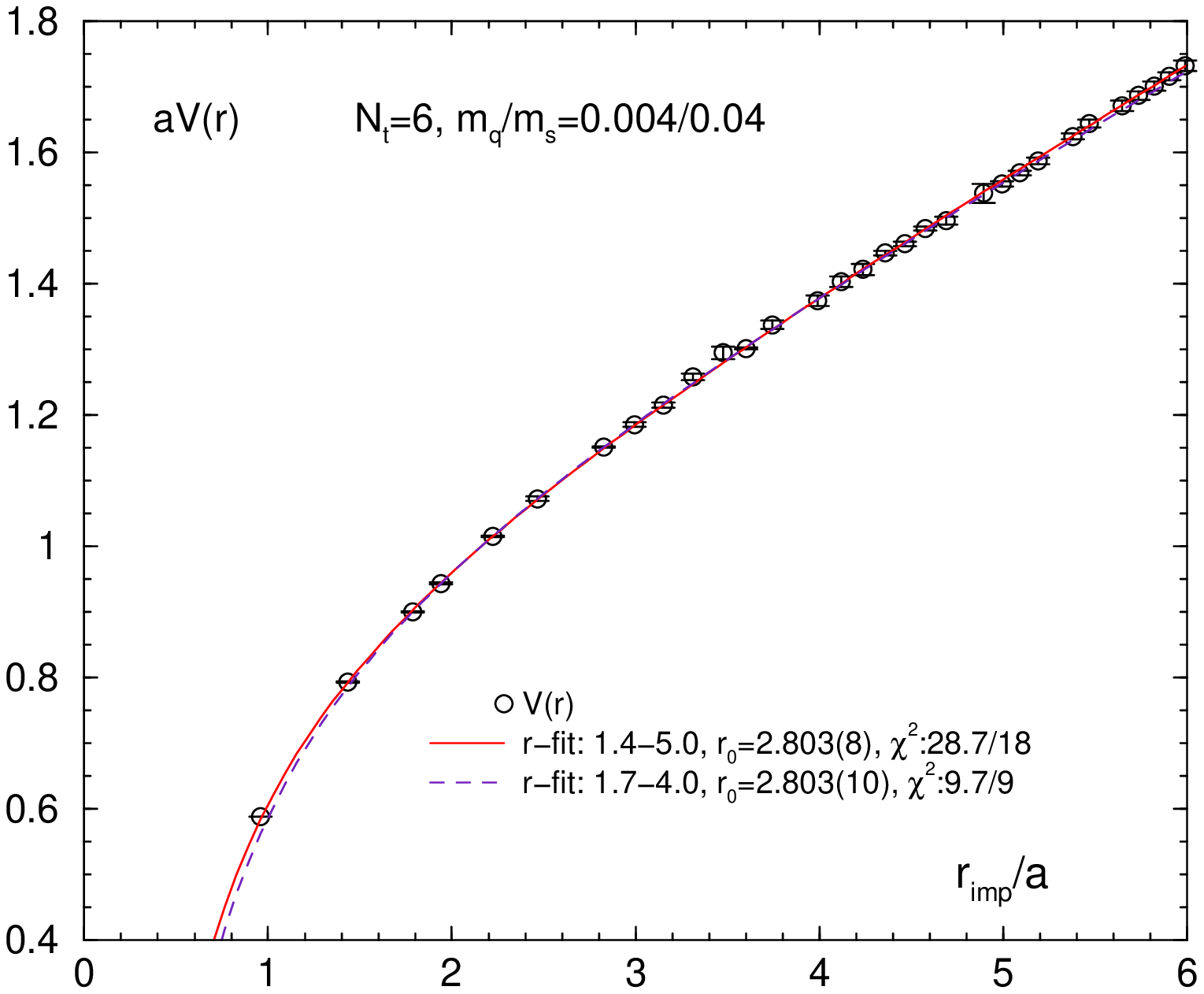}
\includegraphics[height=5.5cm]{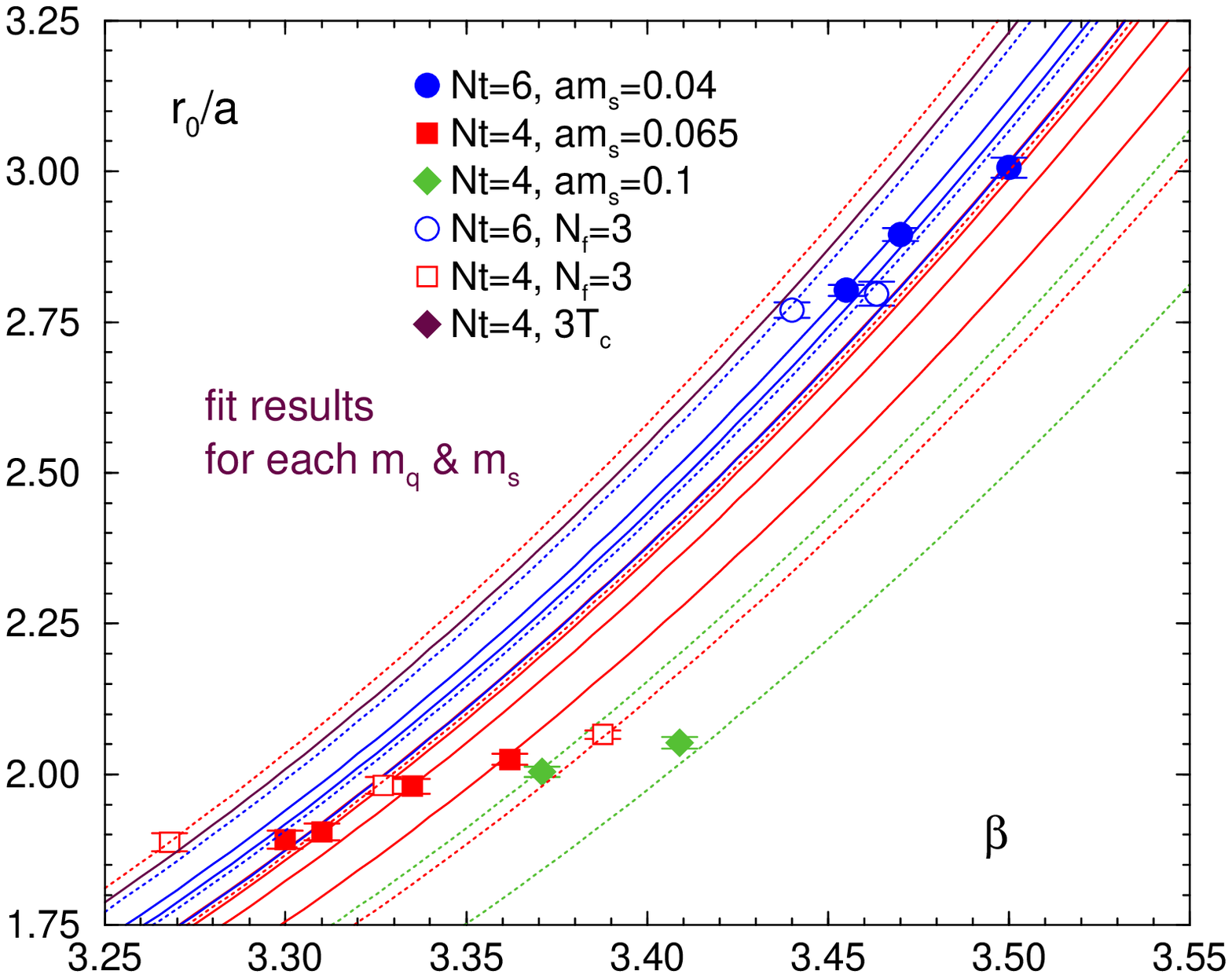}
\end{center}
\end{minipage}
\end{center}
\caption{Fit of the static quark potential with the Cornell ansatz and improved
distances (left) and fit of the scale parameter $r_0$ versus the lattice coupling
$\beta$ to a renormalization group inspired ansatz (Eq~4.3) (right). 
}
\label{fig:hq}
\end{figure}

Our zero temperature calculations have been performed at values of the gauge
coupling in the vicinity of the $\beta_c$ we found from our analysis of the
chiral susceptibilities. We typically generated several thousand configurations
and analyzed the hadron spectrum and static quark potential on every $10^{th}$
configuration. We obtain the scale by using the simple Cornell form to
fit our numerical results for the static quark potential, $V_{\bar{q}q}(r) =
-\alpha/r + \sigma r + c$.  With this fit-ansatz, which does not include a
possible running of the coupling $\alpha$, the force entering the definition of
$r_0$ is easily calculated and we find from Eq.~\ref{r0}, $r_0 \equiv
\sqrt{(1.65 - \alpha)/\sigma}$. 
When fitting the potential, We replace the Euclidean distance 
on the lattice by an improved distance $r_I/a$ which
relates the separation between the static
quark and anti-quark sources to the Fourier transform of the tree-level
lattice gluon propagator.
This procedure removes most of the short distance lattice artifacts.
We show an example for a fit of the static quark potential with improved
distances in Fig.~\ref{fig:hq} (left).

We have determined the scale parameter $r_0$ in units of the lattice
spacing for 9 different parameter sets $(\hm_l, \hm_s, \beta)$. This
allows to interpolate between different values of the gauge coupling
and quark masses.
We use a renormalization group inspired ansatz \cite{allton} which takes
into account the quark mass dependence of $r_0/a$
\cite{Bernard04} and which approaches, in the weak coupling limit,
the 2-loop  $\beta$-function for three massless flavors,
\begin{equation}
(r_0/a)^{-1} =
R(\beta) (1 +B \hat{a}^2(\beta) + C \hat{a}^4(\beta))
{\rm e}^{A (2 \hm_l + \hm_s)+D}
\; .
\label{interpolate}
\end{equation}
Here $R(\beta)$ denotes the 2-loop $\beta$-function
and $\hat{a}(\beta) =R(\beta)/R(\bar{\beta})$ with $\bar{\beta}=3.4$ chosen
as an arbitrary normalization point.
%A fit to 14 values for $r_0/a$, which include 8 of the 9 values for
%$r_0/a$ and additional data obtained in
%our studies of 3-flavor QCD \cite{RHMCtuning}, gives $A=1.45(5)$, $B=1.20(17)$,
%$C=-0.21(6)$ and $D= 2.41(5)$ with a $\chi^2/dof = 0.9$.
We use this interpolation formula to set the scale for the transition
temperature, a fit is shown in Fig.~\ref{fig:hq} (right).

\section{The transition temperature in (2+1)-flavor QCD}
\begin{figure}[tb]
\begin{center}
\begin{minipage}[c]{14.5cm}
\begin{center}
\includegraphics[width=7.2cm]{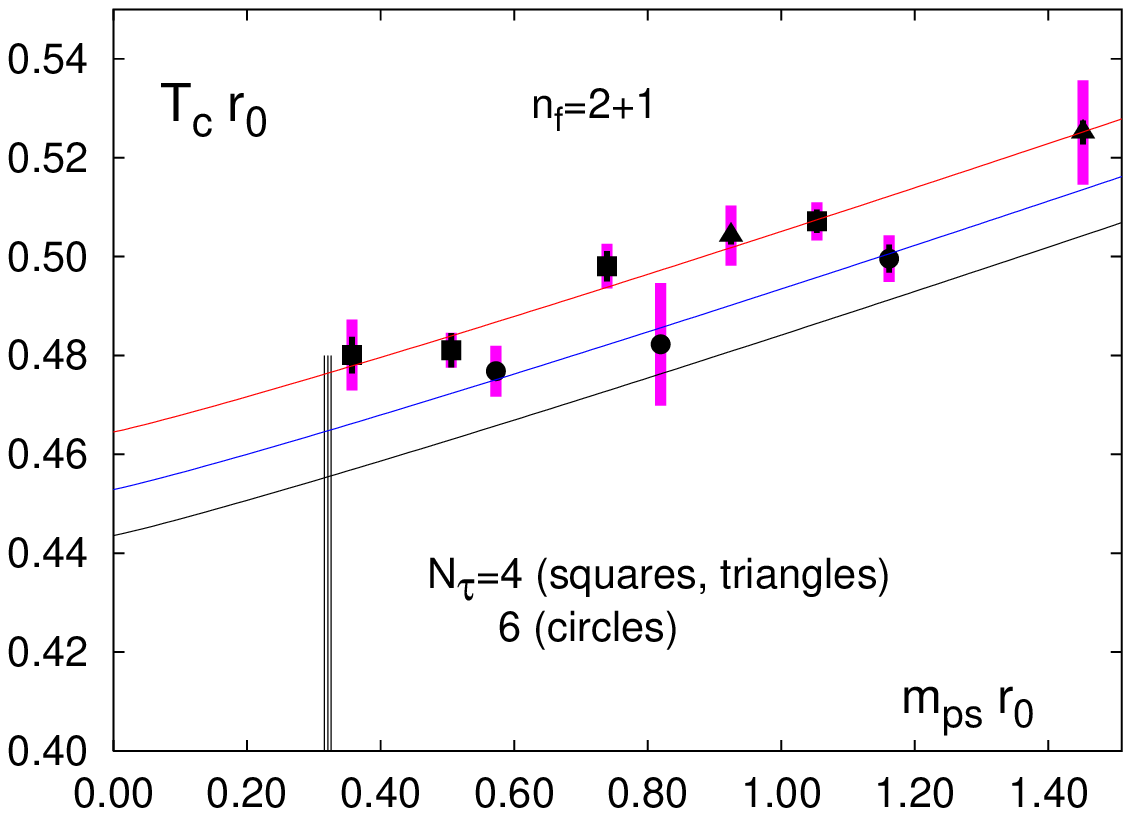}
\includegraphics[width=7.2cm]{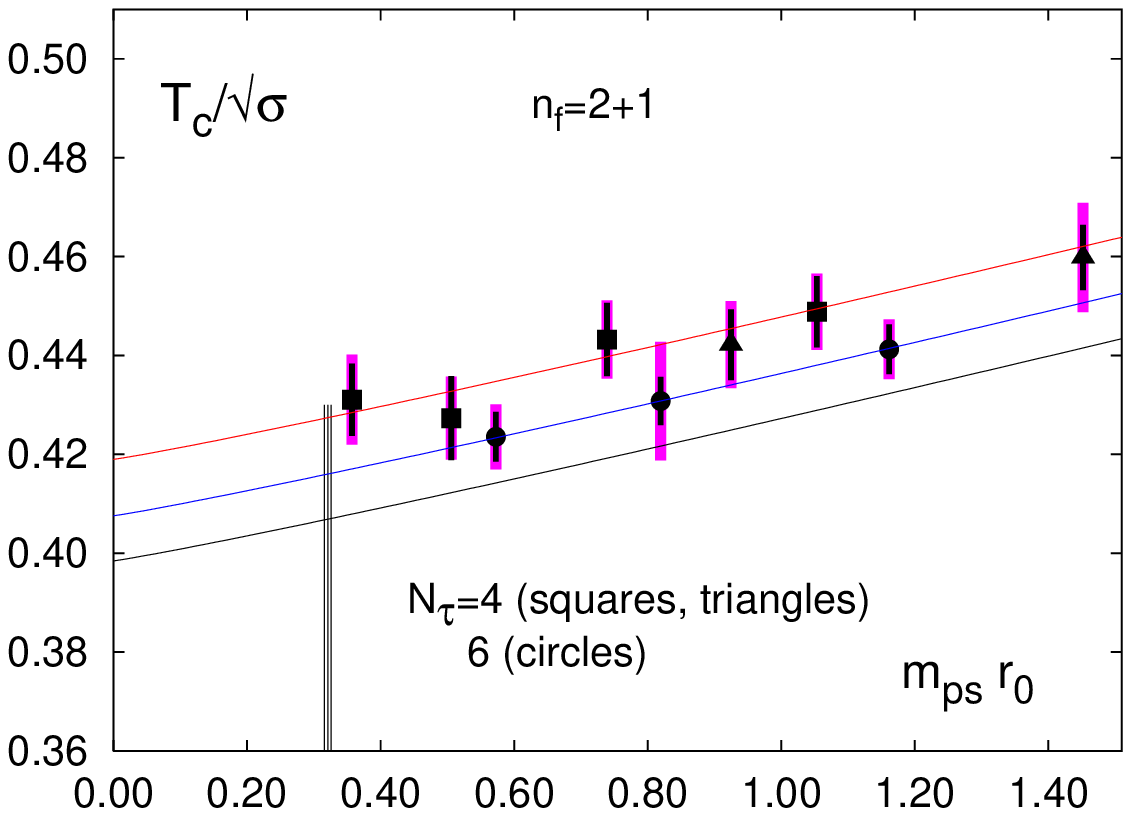}
\end{center}
\end{minipage}
\end{center}
\caption{
$T_c r_0$ (left) and $T_c/\sqrt{\sigma}$ (right) as a function of
$m_{ps}r_0$ on lattices with temporal extent $N_\tau =4$, $\hm_s= 0.065$
(squares) and $\hm_s= 0.1$ (triangles) as well as for
$N_\tau = 6$, $\hm_s= 0.04$ (circles). 
Thin error bars represent the statistical and systematic error on
$r_0/a$ and $\sqrt{\sigma}a$. The broad error bar combines this error with
the error on $\beta_c$.
The vertical line shows the location of the physical value
$m_{ps} r_0= 0.321(5)$ and its width represents the error on $r_0$.
The three parallel lines show results of fits based on Eq.~5.1
with $d=1.08$ for $N_\tau=4,~6$ and $N_\tau \rightarrow \infty$ (top to
bottom).
}
\label{fig:Tc}
\end{figure}

To obtain the transition temperature we use the results for the scales
$r_0/a$ and $\sqrt{\sigma}a$ obtained from fits to the static quark potential.
In cases where zero temperature calculations
have not been performed directly at the critical coupling but at a nearby
$\beta$-value we use Eq.~\ref{interpolate} to determine the scales at
$\beta_c(\hm_l,\hm_s,N_\tau)$.
The transition temperature is then obtained as $T_c r_0 \equiv (r_0/a)/N_\tau$
or $T_c/\sqrt{\sigma}=1/\sqrt{\sigma} aN_\tau$.
We show these results as function of the
pseudo-scalar (pion) mass expressed in units of $r_0$ in Figure~\ref{fig:Tc}.
There we give 2 errors on $T_c r_0$ and $T_c/\sqrt{\sigma}$.
A thin error bar reflects the combined statistical and systematic
errors on the scales $r_0/a$ and $\sqrt{\sigma}a$ obtained from
our 3-parameter fit to the static quark potential.
The broad error bar combines this uncertainty of the zero temperature
scale determination with the scale-uncertainty arising from the error
on $\beta_c$. As can be seen, the former error, which
typically is of the order of 2\%, dominates our uncertainty on $T_cr_0$
and $T_c/\sqrt{\sigma}$
on the coarser $N_\tau=4$ lattices, while the uncertainty in the determination
of $\beta_c$ becomes more relevant for $N_\tau=6$. 

We have extrapolated our numerical results for $T_cr_0$ and
$T_c/\sqrt{\sigma}$,
which have been obtained for a specific set of lattice parameters
($\hm_l, \hm_s, N_\tau$),
to the chiral and continuum limit using an ansatz that takes into account
the quadratic cut-off dependence, $(aT)^2=1/N_\tau^2$, and a quark mass
dependence expressed in terms of the pseudo-scalar meson mass,
\begin{equation}
Y_{ \hm_l,\hm_s,N_\tau } = Y_{ 0,m_s,\infty } +
A (m_{ps}r_0)^d + B/N_\tau^2 \quad,\quad Y=T_cr_0,~T_c/\sqrt{\sigma} \; ,
\label{extrapolation}
\end{equation}
If the QCD transition is second order in the chiral limit
the transition temperature
is expected to depend on the quark mass as $\hm_l^{1/\beta\delta}$, or
correspondingly on the pseudo-scalar meson mass as $m_{ps}^{2/\beta\delta}$
with $d\equiv 2/\beta\delta\simeq 1.08$ characterizing universal scaling
behavior in the vicinity of second order phase transitions belonging to the
universality class of $O(4)$ symmetric, 3-dimensional spin models. If,
however, the transition becomes first order for small quark masses,
which is not ruled out  for physical values of the strange quark mass,
the transition temperature
will depend linearly on the quark mass ($d=2$). A fit to our data set with
$d$ as a free fit parameter would actually favor a value smaller than unity,
although the error on $d$ is large in this case, $d= 0.6(7)$.

Fortunately, the extrapolation to the physical point is not
very sensitive to the choice of $d$ as our calculations have been performed
close to this point. It does, however, increase the uncertainty on
the extrapolation to the chiral limit.
We have performed extrapolations to the chiral limit with $d$
varying between $d=1$ and $d=2$. From this we find

\begin{equation}
m_{ps}r_0 \equiv 0: \qquad T_c r_0 = 0.444(6)^{+12}_{-3} \quad ,
\quad T_c/\sqrt{\sigma} = 0.398(6)^{+10}_{-1} \; ,
\label{Tcchiral}
\end{equation}
where the central value is given for fits with the $O(4)$ exponent
$d=1.08$ and the lower and upper systematic error correspond to
$d=1$ and $d=2$, respectively. Using the fit values for the parameter
$A$ that controls the quark mass dependence of $T_cr_0$ ($A=0.041(5)$)
and $T_c/\sqrt{\sigma}$ ($A=0.029(4)$), respectively, we can determine
the transition temperature at the physical point, fixed by $m_{ps}r_0$r,
where we then obtain a slightly larger value with reduced
systematic errors,
\begin{equation}
m_{ps}r_0 \equiv 0.321(5): \qquad T_c r_0 = 0.457(7)^{+8}_{-2} \quad ,
\quad T_c/\sqrt{\sigma} = 0.408(8)^{+3}_{-1} \; .
\label{Tcphysical}
\end{equation}
Here the error includes the uncertainty in the value for the
physical point, $m_{ps}r_0$, arising from the uncertainty in the scale
parameter $r_0 = 0.469(7)$~fm.
We note that the extrapolated values for $T_c r_0$
and $T_c/\sqrt{\sigma}$ may also be interpreted as a continuum
extrapolation of the shape parameters of the static potential.
This yields  $r_0 \sqrt{\sigma} \simeq 1.11$ which is
consistent with the continuum extrapolation
obtained with the asqtad-action \cite{Bernard04}.

The fit parameter $B$ which controls the size of the cut-off dependent term
in Eq.~\ref{extrapolation} is in all cases close to $1/3$.
We find $B=0.34(9)$ for fits to $T_cr_0$ and $B=0.33(7)$ for fits to
$T_c/\sqrt{\sigma}$, respectively. The critical temperatures for $N_\tau=4$
thus are about 5\% larger than the extrapolated value, and
for $N_\tau=6$ the difference is about 2\%. We therefore expect that
any remaining uncertainties in our extrapolation to the continuum limit
which may arise from higher order corrections in the cut-off dependence
of $T_cr_0$ are not larger than  2\%.

The results for the transition temperature
obtained here for smaller quark masses and smaller lattice spacings is
entirely consistent with the results for 2-flavor QCD obtained previously
with the p4fat3 action on $N_\tau=4$ lattices in the chiral limit,
$T_c/\sqrt{\sigma} =0.425(15)$ \cite{peikert}. We now find for (2+1)-flavor
QCD for $N_\tau=4$ in the chiral limit $T_c/\sqrt{\sigma} = 0.419(6)$.
The continuum extrapolated result is, however, somewhat
larger than the continuum extrapolated result obtained
with the asqtad-action for (2+1)-flavor QCD in the chiral
limit\footnote{In \cite{Bernard04}
$T_c$ is given in units of $r_1$ using results for $r_1/a$ taken from
\cite{MILCpotential}. We have expressed $T_c$ in units of $r_0$
using $r_0/r_1=1.4795$ to convert $r_1$ to the $r_0$ scale used by us.},
$T_c r_0 = 0.402(29)$ \cite{Bernard04}, which is based on
the determination of transition temperatures on lattices with temporal
extent $N_\tau =4$, $6$ and $8$.

Although we frequently have referred to the physical value of $r_0$ during
the discussion in the previous chapters we stress
that our final result for dimensionless quantities, in particular
$T_cr_0$ and $T_c/\sqrt{\sigma}$ given in Eq.~\ref{Tcchiral},
does not depend on the actual physical value of $r_0$ or $\sqrt{\sigma}$.

%Unfortunately neither $r_0$ nor $\sqrt{\sigma}$ are directly measurable
%experimentally. Their physical values have been deduced from lattice
%calculations through a comparison with calculations for the level splitting
%in the bottomonium spectrum \cite{gray,Bernard04}.
%This observable has the advantage of showing only a weak quark mass
%dependence. Of course, dealing with heavy quarks in addition to the
%dynamical light quarks requires a special set-up (NRQCD) which might
%introduce additional systematic errors. However, these findings
%have been cross-checked through calculations of other observables which
%only involve the light quark sector. In particular, the
%pion decay constant, $f_\pi$,
%has been evaluated on the MILC configurations that have
%been used for the bottomonium level splitting and yields a consistent
%value for $r_0$ \cite{Davies}.
%One may, of course, also  consider using results for masses of mesons
%constructed from light quarks, e.g. the vector meson mass, to determine the
%scale for the transition temperature, eg. $T_c/m_\rho$. However,
%even on lattices with smaller lattice spacings than those used in
%thermodynamic calculations today, the calculation of vector meson
%mass is known to suffer from large statistical and systematic
%errors \cite{MILCpotential,Davies}.
%This is even more the case on the coarse lattices needed
%for our finite temperature calculations.
%We thus refrained from using results on the vector meson mass for
%our determination of the transition temperature.

At present the scale parameter $r_0$, deduced from the bottomonium
level splitting using NRQCD \cite{gray}, seems to be the best controlled
lattice observable that can be used to set the scale for $T_c$.
Using for $r_0$ the value given in Eq.~\ref{r0_fm} we obtain for the transition
temperature in QCD at the physical point,
\begin{equation}
T_c = 192(7)(4)\; {\rm MeV} \; ,
\label{TcMeV}
\end{equation}
where the statistical error includes the errors given in Eq.~\ref{Tcphysical}
as well as the uncertainty in the value of $r_0$ and the second error reflects
our estimate of a remaining systematic error on the extrapolation to the
continuum limit. As discussed after Eq.~\ref{Tcphysical} we estimate this error
which arises from neglecting higher order cut-off effects in our ansatz
for the continuum extrapolation, Eq.~\ref{extrapolation} to be about 2\%.
Our resut for $T_c$ agrees within our current uncertainties with the result by
the MILC Collaboration which is $T_c=169(12)(4)$ \cite{Bernard04}. However, we
encounter a discrepancy with the result by the Wuppertal group of
$T_c=151(3)(3)$ \cite{fodor:tc} as determined by the chiral condensate.

%The value of the critical temperature obtained here is about 10\% larger
%than the frequently quoted value $\sim 175$~MeV.
%We note that this larger value mainly results from
%the value for $r_0$ used in our conversion to
%physical scales. Together with $r_0\sqrt{\sigma} \simeq 1.11$
%it implies that the string tension takes on
%the value $\sqrt{\sigma} \simeq 465$~MeV. This value of the string tension
%is about 10\% larger than that used in the past to set the scale for $T_c$
%\cite{peikert}.

\section{The equation of state}
We calculate the equation of state on a line of constant physics (LCP). Along
this line the physical quark masses are fixed but the bare light and strange
quark masses, which are the parameters in our action, change as a function of
the coupling. We define the LCP by demanding that the ratio of strange
pseudo-scalar mass over kaon mass ($m_{\bar{s}s}/m_K$) as well as
$m_{\bar{s}s}$ in units of the scale parameter ($r_0 m_{\bar{s}s}$) stay constant. We
find that in good approximation the first condition is fulfilled by holding the
ratio of bare light and strange quark mass ($\hat{m}_l/\hat{m}_s$) fixed. We
calculate meson masses in a wide range of the parameter space and use
a renormalization group inspired interpolation formula to carefully tune the
light quark masses as a function of the coupling $\beta$ in order to satisfy
the second condition.  

For calculating the pressure and the interaction measure we employ the
integral method. Since the logarithm of the partition function can not be
calculated easily on the lattice,  derivatives with respect to the bare 
parameters are calculated ($\partial \ln Z/\partial \beta$, $\partial \ln
Z/\partial \hat{m}_l$, $\partial \ln Z/\partial \hat{m}_s$) and integrated 
along the LCP. We find for the pressure
\begin{eqnarray}
\left.\frac{p}{T^4}\right|_{\beta_0}^{\beta}&=&N_\tau^4\int_{\beta_0}^{\beta}d\beta'\;
\Bigg[\frac{1}{N_\sigma^3 N_\tau}\left(\left<S_G\right>_0-\left<S_G\right>_T\right)\nonumber \\
&&-\left(2\left(\left<\bar\psi\psi\right>_{l0}-\left<\bar\psi\psi\right>_{lT}\right)
  +\frac{\hat{m}_s}{\hat{m}_l}
   \left(\left<\bar\psi\psi\right>_{s0}-\left<\bar\psi\psi\right>_{sT}\right)
   \right)\left(\frac{\partial \hat{m}_l}{\partial\beta'}\right)_{\hat{m}_s/\hat{m}_l}\nonumber \\
&&-\hat{m}_l\left(\left<\bar\psi\psi\right>_{s0}-\left<\bar\psi\psi\right>_{sT}\right)
   \left(\frac{\partial \hat{m}_s/\hat{m}_l}{\partial\beta'}\right)_{\hat{m}_l} \Bigg]\quad .
\end{eqnarray}
As we find that $\hat{m}_s/\hat{m}_l=10$ can be kept fixed on a LCP, the last
term in the integral vanishes. To calculate the interaction measure $\varepsilon-3p$, the
lattice $\beta$-function has to be known. Here we have
\begin{eqnarray}
\frac{\varepsilon-3p}{T^4}
&=&T\frac{{\rm d}}{{\rm d}T}\left(\frac{p}{T^4}\right)
  =-\left(a\frac{{\rm d}\beta}{{\rm d}a}\right)_{\rm LCP}
    \frac{{\rm d}p/T^4}{{\rm d}\beta}\nonumber \\
&=&\left(\frac{\varepsilon-3p}{T^4}\right)_{\rm gluon}
  +\left(\frac{\varepsilon-3p}{T^4}\right)_{\rm fermion}
  +\left(\frac{\varepsilon-3p}{T^4}\right)_{\hat{m}_s/\hat{m}_l}\quad .
\end{eqnarray}
Again the last term vanishes to good accuracy on the LCP. We emphasize that a
precise knowledge of the $\beta$-function, $a{\rm d}\beta/{\rm d}a$, along the
LCP is necessary to calculate the interaction measure and energy density. We
calculate the $\beta$-function from measurements of the scale parameter $r_0$,
which we perform on each temperature (coupling) which we include in our
equation of state. 

\begin{figure}
\begin{center}
\includegraphics[width=7cm]{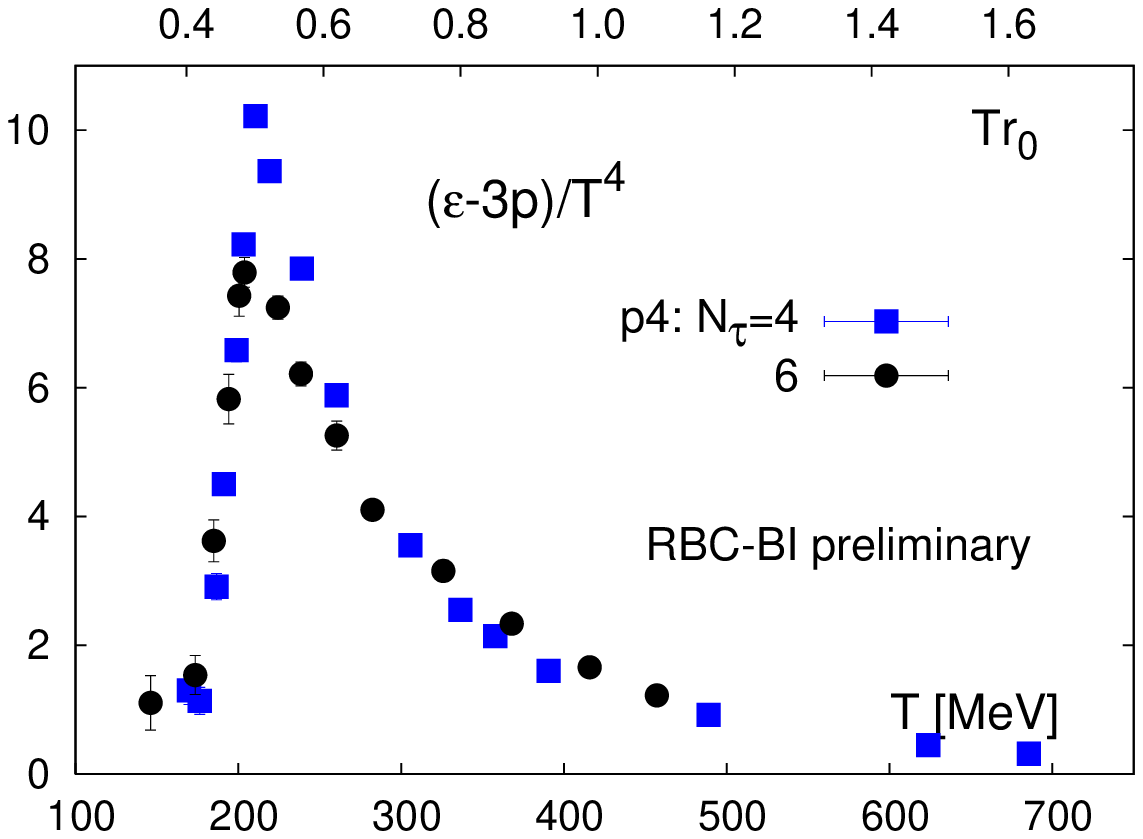}
\includegraphics[width=7cm]{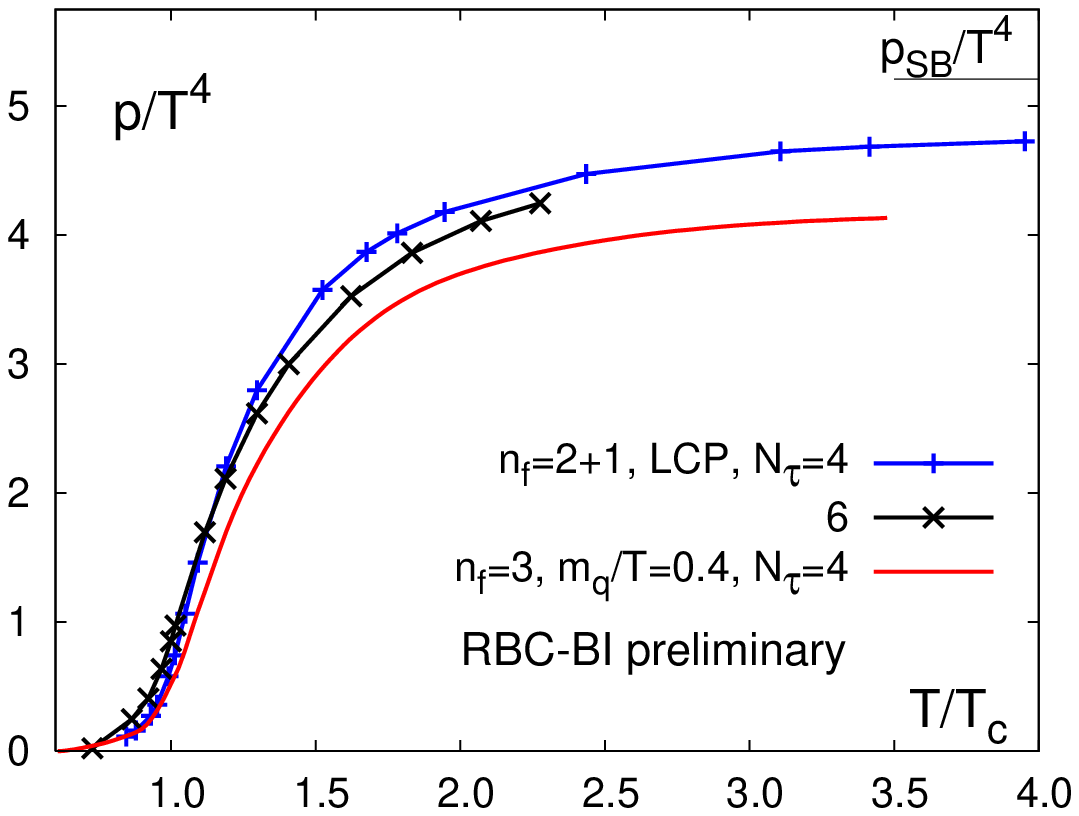}
\end{center}
\caption{The interaction measure (left) and pressure (right) along a
  line of constant physics for $N_\tau=4$ and 6. The ratio of light to strange
  quark mass has been held fixed at $\hat{m}_l/\hat{m}_s=0.1$. Also shown are earlier
  results for the pressure, obtained with 3-flavor QCD and const. $m_q/T=0.4$ [2].}
\label{fig:eos}
\end{figure}

In Fig.~\ref{fig:eos} we show our preliminary results for the interaction
measure (left) and the pressure (right) on the LCP for $N_\tau=4$ and
6. The ratio of light to strange quark mass has been held fixed at
$\hat{m}_l/\hat{m}_s=0.1$. 
We compare our result for the pressure with earlier results for 3-flavor QCD
and a constant $m_q/T=0.4$ \cite{peikert_pressure}. One clearly sees a mass
dependence, however, the mass dependence is small. Note that the quark masses differ by
more than an order of magnitude in the high temperature region.

Or results on the interaction measure and pressure obtained with the p4fat3
improved action are in complete agreement
with corresponding results obtained with asqtad fermions \cite{Bernard:2006nj}.
Furthermore, the results show little cut-off dependence in the entire
temperature regime analyzed. This is in contrast to calculations performed with
the standard staggered discretization scheme \cite{fodor:eos}, which leads to
large cut-off effects in the hight temperature limit for lattices with temporal
extent of $N_\tau=4$ and 6.

%We note that we find the increase in the energy
%density coincides with our estimate of the critical temperature of 192(7)~MeV
%although the absolute temperature scale does not depend on the
%$T_c$ determination. Furthermore, it is interesting to remark that the value of
%the energy density at $T_c$ ($\varepsilon_c/T^4\approx 6$) did almost not change
%from previous calculations \cite{peikert_pressure},i.e. the critical energy
%density depend only little on the quark mass. 

\section{Hadronic fluctuations at zero and non-zero density}
\label{sec:fluct}

It is conceptually very simple to calculate the expansion coefficients
of any observable $O$ in a Taylor series around $\mu_q=0$ where $\mu_q$ is the
quark chemical potential:
\begin{equation}
O(\hat{\mu}) = c_0+c_1\hat{\mu}+\frac{1}{2}c_2\hat{\mu}^2+\cdots \quad .
\label{taylor_series_O}
\end{equation}
Since on the lattice all quantities are given in units of the lattice spacing
($a$), the expansion parameter is $\hat{\mu}\equiv a\mu_q=N_\tau^{-1}(\mu_q/T)$.
This idea goes back to the first calculation of the quark number susceptibility
\cite{Gottlieb:1988cq}. The response of hadron masses \cite{Choe:2001ar} as
well as the pressure and further bulk thermodynamic quantities
\cite{Gavai:2003mf,Allton:2003vx,Allton:2005gk,Ejiri:2005uv} have been studied
by this method. The first two nontrivial coefficients in
Eq.~(\ref{taylor_series_O}) are given by
\begin{eqnarray}\label{taylor_coeff_O}
c_1&=&  \left< \frac{\partial O           }{\partial \hat{\mu}}\right>
      \!+\! \left<O\frac{\partial \ln\rm{det}D}{\partial \hat{\mu}}\right>\\
c_2&=&  \left< \frac{\partial^2 O}{\partial \hat{\mu}^2}\right>
      \!+\!2\left< \frac{\partial O}{\partial \hat{\mu}}
               \frac{\partial \ln\rm{det}D}{\partial \hat{\mu}}\right>
      \!+\! \left<O\frac{\partial^2 \ln\rm{det}D}{\partial \hat{\mu}^2}\right>
      \!-\! \bigg<O\bigg>\left<\frac{\partial^2 \ln\rm{det}D}{\partial
      \hat{\mu}^2}\right> \quad . \nonumber
\end{eqnarray}
Besides derivatives of the observable itself, the calculation of
derivatives of $\ln \rm{det}D$ with respect to $\hat{\mu}$ is required.
The derivatives have to be taken at $\hat{\mu}_0=0$. Note that due to the
particle-antiparticle 
symmetry of the partition function ($Z(\mu_q)=Z(-\mu_q))$ all odd
coefficients in Eq.~(\ref{taylor_series_O})  vanish identically. For the
same reason we have
$\left<\partial \ln\rm{det} D/\partial \hat{\mu}\right>=0$ at
$\hat{\mu}=0$. We explicitly use this property in
Eq.~(\ref{taylor_coeff_O}) to derive the expansion coefficients.

The advantages of this method are that expectations values only have to be
evaluated at $\hat{\mu}=0$, i.e. calculations are not directly affected by
the sign problem.  Furthermore, all derivatives of the fermion determinant
can be expressed in terms of traces by using the identity
$\ln\rm{det}D=\rm{Tr}\ln D$. This enables the stochastic calculation of the
expansion coefficients by the random noise method, which is much faster than
a direct evaluation of the determinant. Moreover, the continuum and infinite
volume extrapolations are well defined on a coefficient by coefficient basis.

Quark number fluctuations $\chi_q$ belong to the most important observables
that allow to follow the transition line into the non-zero chemical potential
plane. They diverge at the critical end-point and thus provide an excellent
signal for its existence and location. Eventually these fluctuations may
be detectable in heavy ion experiments. Hadronic fluctuations can be computed
from Taylor expansion coefficients of the pressure with respect to the quark
chemical potential:
\begin{equation}
\frac{p}{T^4}
%\equiv \Omega(T,\mu_q)
= \sum_{n=0}^{\infty} c_n(T)
\left(\frac{\mu_q}{T}\right)^n \quad \mbox{with} \quad
c_n(T) =
%\left. \frac{1}{n!}\frac{\partial^n\Omega}{\partial(\mu_q/T)^n}
%\right|_{\mu_q=0}=
\left. \frac{1}{n!}\frac{N_\tau^3}{N_s^3}\frac{\partial {\rm ln}
Z}{\partial(\hat\mu N_\tau)^n} \right|_{\hat\mu=0} \quad .
\end{equation}
Note that in the Taylor expansion of the pressure the up and down quark
chemical potentials have been chosen to be equal. Having calculated the
coefficients $c_n(T)$ one can construct the quark number density and quark
number fluctuations
\begin{equation}
\frac{n_q}{T^3}=\sum_{n=2}^{\infty}nc_n(T)\left(\frac{\mu_q}{T}\right)^{n-1}
\quad ; \quad
\frac{\chi_q}{T^2}=\sum_{n=2}^{\infty}n(n-1)c_n(T)
\left(\frac{\mu_q}{T}\right)^{n-2} \quad .
\label{Eq:chiq}
\end{equation}

\begin{figure}
\begin{center}
\begin{minipage}[c]{0.329\textwidth}
\includegraphics[width=1.0\textwidth]{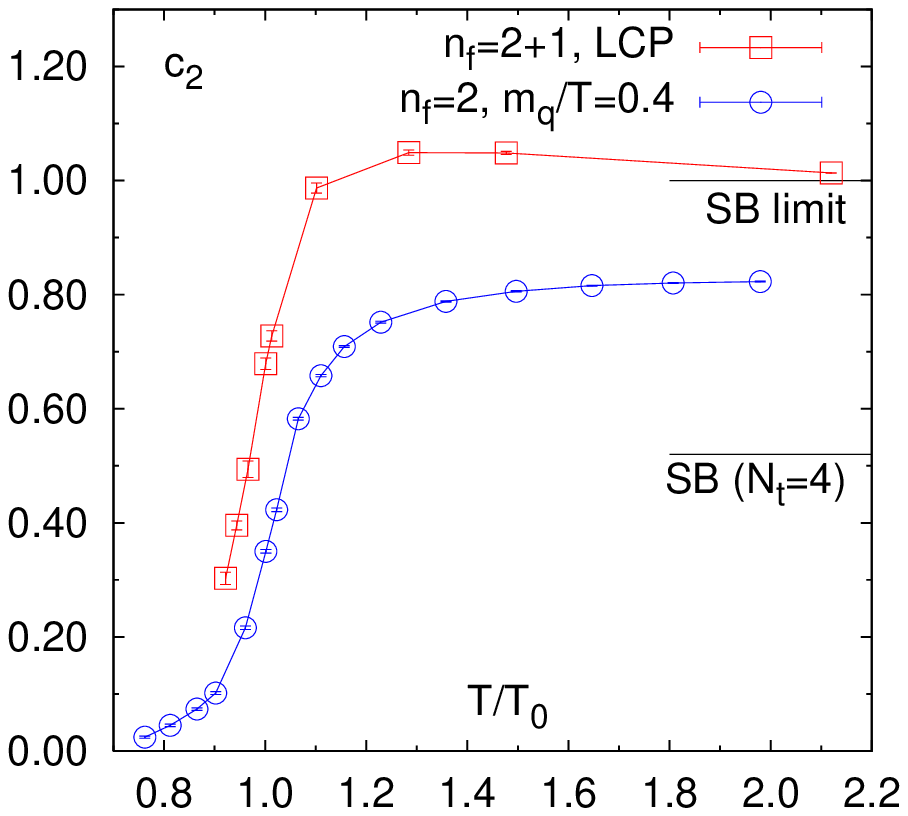}
\end{minipage}
\begin{minipage}[c]{0.329\textwidth}
\includegraphics[width=1.0\textwidth]{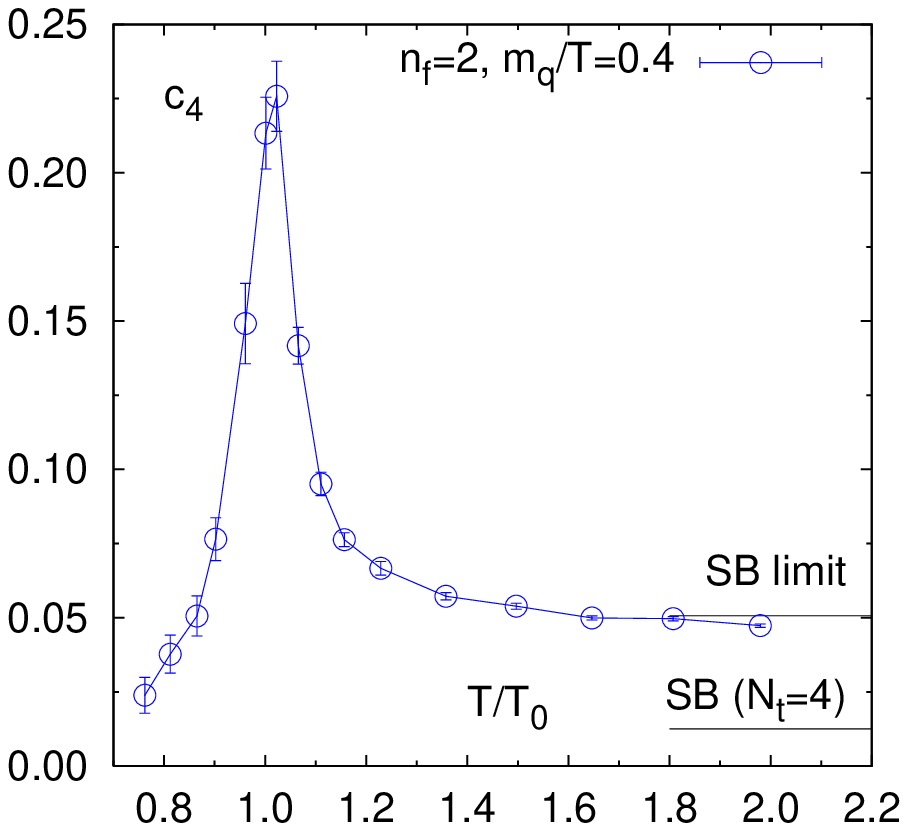}
\end{minipage}
\begin{minipage}[c]{0.329\textwidth}
\includegraphics[width=1.0\textwidth]{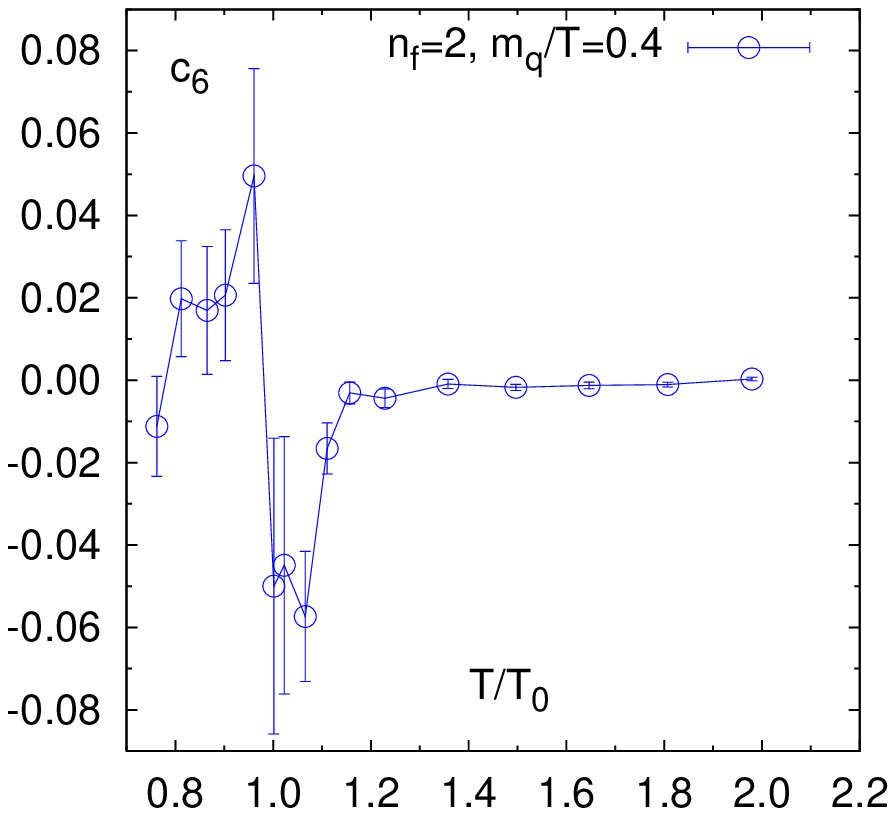}
\end{minipage}
\end{center}
\caption{The Taylor expansion coefficients $c_2$, $c_4$ and $c_6$ of the
pressure for $n_f=2$ and $m_q/T=0.4$ [27]. Also shown are preliminary results
for (2+1)-flavor QCD and a quark mass ratio of $\hat{m}_q/\hat{m}_s=0.1$.}
\label{fig:c2c4c6}
\end{figure}

In the case of two flavors of p4-improved staggered fermions, 
with $m_q/T=0.4$ the first three non-zero coefficients 
$c_2$, $c_4$, and $c_6$ have been calculated \cite{Allton:2005gk} and are shown
in Fig.~\ref{fig:c2c4c6}. We also show our preliminary results on $c_2$ for
(2+1)-flavor QCD and a quark mass ratio of $\hat{m}_q/\hat{m}_s=0.1$. A
mass dependence is clearly evident.

\begin{figure}
\begin{center}
\begin{minipage}[c]{0.45\textwidth}
\includegraphics[width=1.0\textwidth]{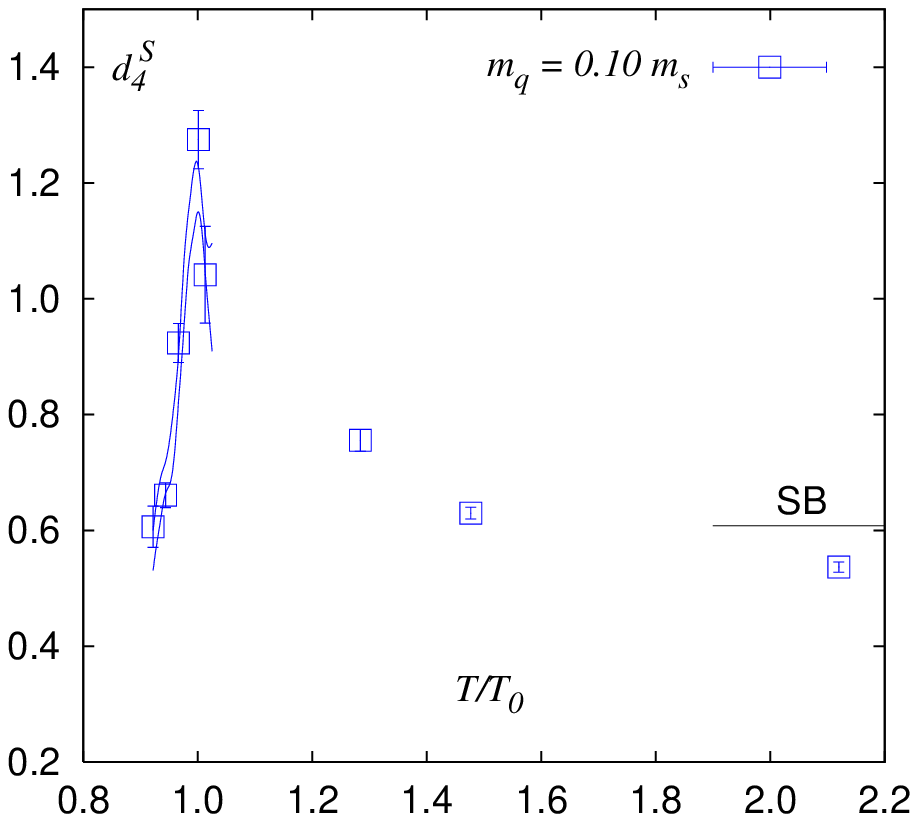}
\end{minipage}
\begin{minipage}[c]{0.45\textwidth}
\includegraphics[width=1.0\textwidth]{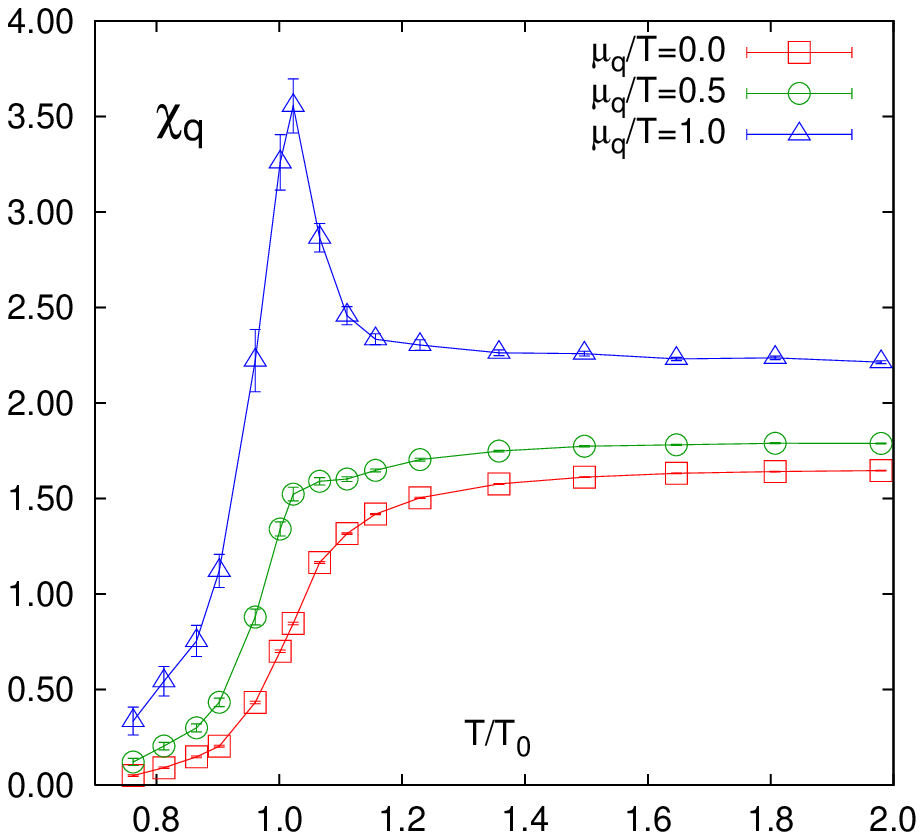}
\end{minipage}
\end{center}
\caption{Quartic strange quark number fluctuations for
  (2+1)-flavor QCD and a quark mass ratio of $\hat{m}_q/\hat{m}_s=0.1$ (left)
  and the quark number susceptibility for $n_f=2$ and
  $m_q/T=0.4$ at several values of the quark chemical potential (right).}
\label{fig:d4}
\end{figure}

In Fig.~\ref{fig:d4} (left) we show the quartic quark number fluctuations of 
strange quarks from our ongoing (2+1)-flavor QCD simulations. 
 which are given by $d_4^S \equiv (1/(VT^3))(\partial^4 \ln Z/\partial (\mu_s/T)^4)$. 
As one can see also the quartic strange quark fluctuations show a peak at
$T_c$. 

In Fig.~\ref{fig:d4}(right) we show the quark
number fluctuations for $n_f=2$ and $m_q/T=0.4$ \cite{Allton:2005gk} for
several values of the quark chemical potential including only the leading
$(mu_q/T)^2$ correction which is proportional to $c_4$ . It is interesting to
see that at $\mu_q=0$, $\chi_q$ shows a rapid but monotonic increase at 
the transition temperature, whereas a cusp is developing at $T_c(\mu_q)$ for
$\mu_q>0$. This is a clear sign for approaching the critical end-point. 

\section{The critical end-point}
\label{sec:EP}

Locating the critical point is one of the most challenging goals of lattice
QCD calculations at finite chemical potential. The first attempt to locate the critical
point used the reweighting method \cite{Fodor:2001pe}. For this calculation,
2+1 flavor of standard staggered fermions have been used at a pion mass of
about 300 MeV and a kaon mass of about 500 MeV. Lattice sizes have, however,
been rather small ($4^3\times 4$ - $8^3\times 4$). A critical chemical
potential of $\mu_B^{crit}=725(35)$ MeV was found. A second calculation
\cite{Fodor:2004nz}, using again the reweighting method, with physical masses
($m_\pi=150$ MeV, $m_K=500$ MeV) and somewhat larger volume
($6^3\times 4$ - $12^3\times 4$), let to $\mu_B^{crit}=360(4)$ MeV.

When using the reweighting method for locating the critical point, the
minima of the normalized partition function in the complex $\beta$-plane
(Lee-Yang zeros) have to be determined
\begin{equation}
Z_{\rm norm}\equiv \left| \frac{ Z(\beta_{\rm Re}, \beta_{\rm Im},\mu) }{
Z(\beta_{\rm Re}, 0,0)} \right|
=
\left| \left< e^{6i \beta N_\tau N_\sigma^3 \Delta S_G} e^{i\theta}
e^{( N_f/4)( {\rm ln det} D(\mu)
- {\rm ln det} D(0) ) }\right>_{(\beta_{\rm Re},0,0)} \right|\quad.
\end{equation}
In $SU(3)$ gauge theory, where we have $e^{i\theta}=1$, this can be done
with high accuracy \cite{Ejiri:2005ts}.
In QCD with non-zero chemical potential the analysis of
Lee-Yang zeros is, however, subtle \cite{Ejiri:2005ts}. For large volumes and
chemical potentials the phase factor of the determinant $e^{i\theta}$ will
force the Lee-Yang zero onto the real axis, which might lead to an
underestimation of the critical point.

Another difficulty with the reweighting method at finite chemical potential
has been pointed out in \cite{Golterman:2006rw}. It was noted, that taking the
fourth (or square) root of the determinant (which is necessary in order to
simulate 2 or 1-flavor QCD with staggered fermions; see also \cite{Sharpe})
could lead to phase ambiguities. This problem becomes acute when
$\mu_q>m_\pi/2$.

All of the above mentioned limitations are, however, irrelevant for the
location of the critical point with the reweighting method if the critical
point is located at small values of $\mu_q$.

Using the Taylor expansion coefficients of the pressure, it is also possible
to estimate the location of the critical point. The convergence radius of the
expansion is limited by the nearest singularity in the complex chemical
potential plane. For each fixed temperature, the radius of convergence is
given by
\begin{equation}
\rho=\lim_{n\to\infty}\rho_n=\lim_{n\to\infty}\sqrt{\left|
\frac{c_n}{c_{n+2}}\right|}\quad.
\end{equation}
Moreover, the sign of the coefficients $c_n$ gives information about the
location of the singularity in the complex plane. If all coefficients are
positive, the singularity is located on the real axis of the complex chemical
potential plane. If the sign is strictly alternating, the singularity lies on
the imaginary axis. For a detailed discussion see \cite{Stephanov:2006dn}.

\begin{figure}
\begin{center}
\begin{minipage}{0.4\textwidth}
\includegraphics[height=5.0cm]{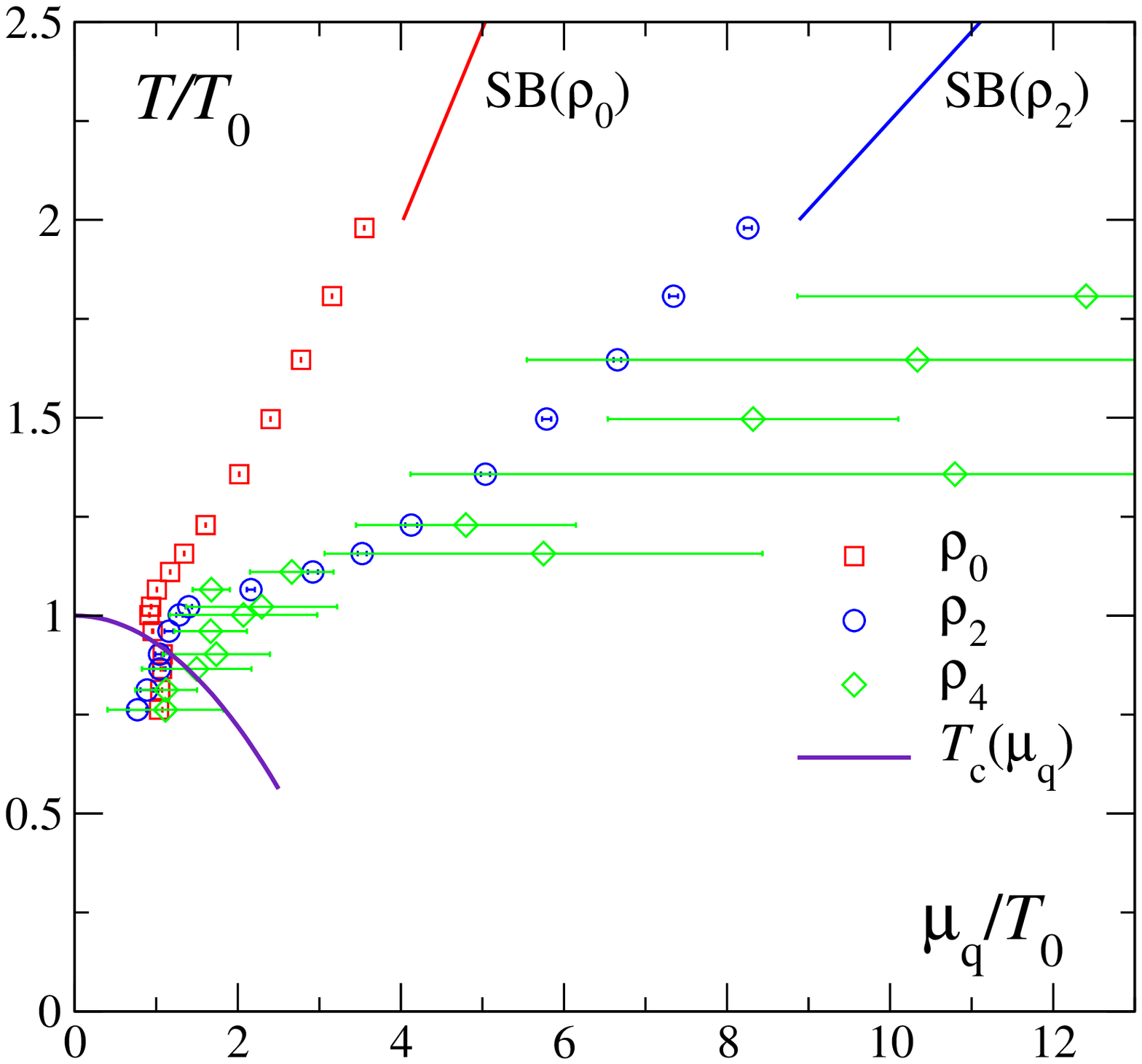}
\end{minipage}
%\hfill
\begin{minipage}{0.5\textwidth}
\includegraphics[width=1.0\textwidth]{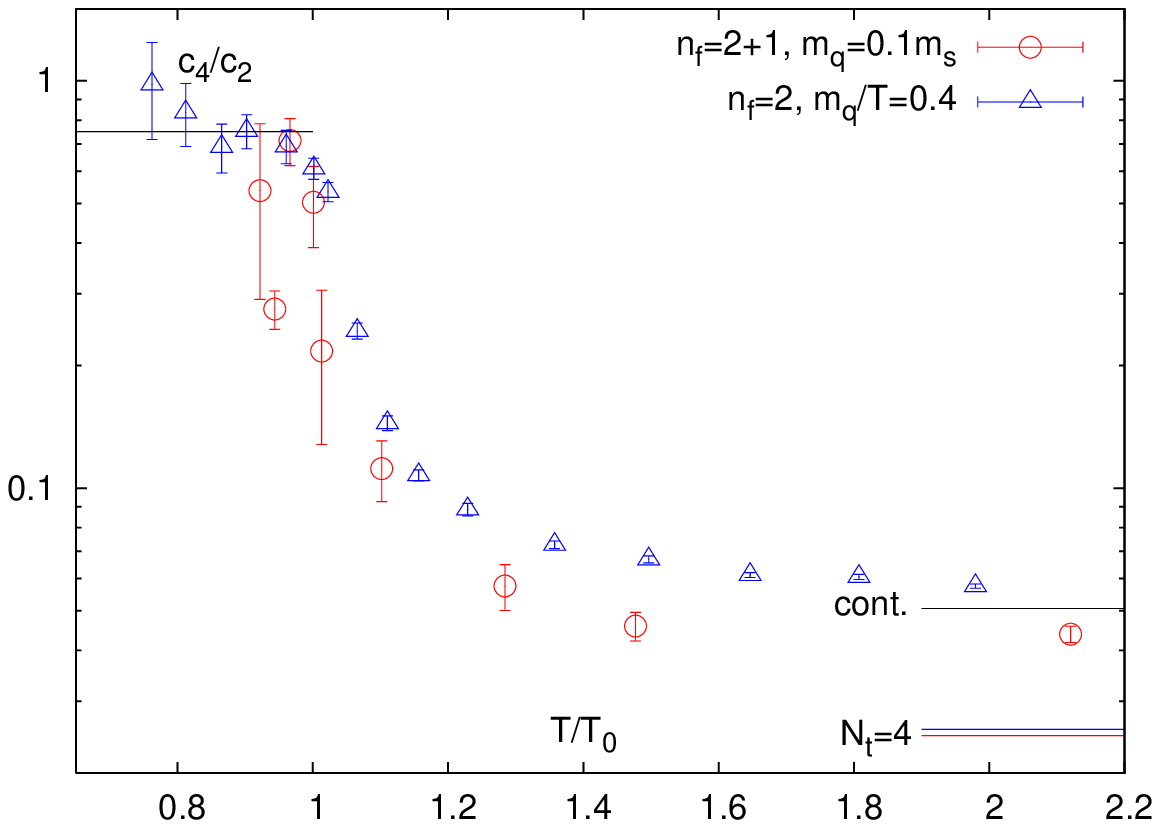}
\end{minipage}
\end{center}
\caption{Estimates of the radius of convergence in the $(T,\mu_q)$-plane
(left), the ratio $c_4/c_2$ of the expansion coefficients (right).
The horizontal lines indicate the resonance gas limit ($T\to 0)$ and the SB
limit in the continuum and at $N_\tau=4$ ($T\to\infty$).
\label{fig:rho}}
\end{figure}

Having only a limited number of expansion coefficients, one can only estimate
$\rho$. The hope is that the convergence of the $\rho_n$ will be fast. Indeed,
a clustering of the $\rho_n$ is seen in the phase diagram, as shown in
Fig.~\ref{fig:rho} \cite{Allton:2005gk}.
This calculation, which has been performed with 2 flavors of p4 improved
fermions and $m_\pi/m_\rho=0.7$, suggests a critical chemical potential of
$\mu_B^{crit}\approx 500$~MeV. All calculated $\rho_n$ are, however,
consistent (within statistical error) with the resonance gas model in the
Boltzmann approximation, where the radius of convergence is infinity.

The authors of \cite{Gavai:2004sd} have estimated the critical chemical
potential from a Taylor expansion of the quark number susceptibility and
find $\mu_B^{crit}\approx 180$~MeV. Two flavors of standard staggered fermions
have been used on lattices up to $24^3\times 4$ and quark mass corresponding
to $m_\pi/m_\rho=0.3$. The difference between the two estimates
\cite{Allton:2005gk,Gavai:2004sd} of the critical point is large. We note
that the second estimate comes from the expansion coefficients of $\chi_q$.
As can be seen from Eq.~\ref{Eq:chiq} this will result in a smaller $\rho_n$
for each fixed $n$. The limit $\lim_{n\to\infty}\rho_n$ is of course the same.
For finite $n$, however, the estimate of $\mu_B^{crit}\approx 180$ MeV would
correspond to $\mu_B^{crit}\approx 240$ MeV, when estimating the $\rho_n$ with
coefficients of the same order from the expansion of the pressure. Nonetheless,
the difference between the two estimates is still striking. The origin
could be the difference in mass. However, preliminary results from the
RBC-Bielefeld Collaboration, also shown in Fig.~\ref{fig:rho}, do not indicate
a strong mass dependence in $c_4/c_2=1/\rho_2^2$.

\section{Beyond the critical point}
\label{sec:beyond}
\begin{figure}
\begin{center}
\includegraphics[width=.5\textwidth]{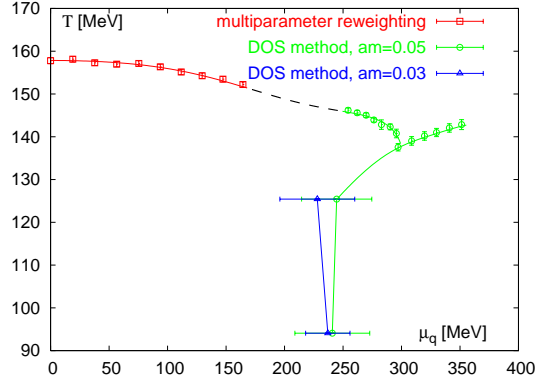}
\caption{Phase diagram from the density of state method [35].\label{fig:dos}}
\end{center}
\end{figure}

Even more challenging than locating the critical point, is the study of the
physics at high densities and low temperatures. One attempt to do so is a
calculation using the density of states method \cite{dos}. Using four flavors
of standard staggered fermions (i.e. taking the root of the determinant is not
necessary), several simulation points in the $(\beta,\hat\mu)$ plane have been
chosen to generate phase quenched configurations by employing the method
proposed in \cite{Kogut:2002zg}. The lattice size has been $6^3\times 4$, $6^4$
and $6^3\times 8$. The quark mass was chosen to be $m/T=0.3$. The generation
has been done with constrained plaquettes. In oder to do so, we introduce a
sharply peaked Gaussian potential in to the partition function, which in
practice leads to a modification of the force term of the HMD-R algorithm. For
each simulation point, several runs have been performed with about 20 different
values of the plaquette. By calculating the eigenvalues of the reduced matrix
the phase of the determinant was calculated for each of those runs. By
numerically calculating the integrals
\begin{equation}
\left<P\right>=\int dx\; x\rho(x)\left<{\rm cos}(\theta)\right>_x \quad
\left<P^2\right>=\int dx\; x^2\rho(x)\left<{\rm cos}(\theta)\right>_x \quad,
\end{equation}
we recover the grand canonical expectation value of the plaquette and its
square. Here $\rho(x)$ is the density of states, which has
been measured by the integral method, usually used to calculate the pressure.
The susceptibility of the plaquette is then given by the usual expression
$\chi_P=\left<P^2\right>-\left<P\right>^2$. From the peak position of the
plaquette susceptibility the phase diagram was calculated as shown in
Fig.~\ref{fig:dos}.
The points at $T=93~\mbox{MeV}$ are from calculations on $6^3\times 8$
lattices. Note, that we make no statement about the order of the
transition lines. To determine the order of the transition one has to perform
a finite-size-scaling analysis which is beyond the scope of this article.

The plaquette expectation value and plaquette susceptibility suggest three
different phases, which coincide in a triple point. The triple point is located
around $\mu_q^{\rm tri}\approx 300~\mbox{MeV}$, 
however its temperature ($T^{\rm tri}$) decreases from $T^{\rm tri}\approx
148~\mbox{MeV}$ on the $4^4$ lattice to $T^{\rm tri}\approx 137~\mbox{MeV}$ on
the $6^4$ lattice. This shift reflects the relatively large cut-off effects one
faces, with standard staggered fermions and temporal extents of 4 and 6. 

The new phase at large chemical potentials and low temperatures is a
natural candidate for a color superconducting phase. 
Recently, by combining experimental results from cold atoms in a trap \cite{atoms}
and some universal arguments, an upper bound for the transition line from the
quark gluon plasma phase (QGP) to the superconducting phase (SC) was proposed
($T_c\le0.35 E_F$) \cite{Schaefer, Shuryak}. To first approximation the Fermi-Energy
$E_F$ is given by the chemical potential $\mu_q$. In \cite{Shuryak} the triple point was
estimated by comparing this upper bound with the experimental freeze-out curve.
A temperature of $T^{\rm tri}\le70~\mbox{MeV}$ was found.
%In Figure~\ref{fig:dos} the upper bound from the experiment is shown as
%solid black line. Our continuum extrapolation of the QGP/SC transition line,
%shown as doted line, yield a larger upper bound of $T_c\le0.45 E_F$.
%Since cold atoms are not quarks, the two estimates do not have to agree. 
Our value of the triple-point roughtly corresponds to $T_c\le0.46 E_F$.
It is interesting that the two values are close.

At low temperatures we find a phase boundary which is very steep
and almost independent of $\mu_q$. Although our lowest temperature is 
$96~\mbox{MeV}$ an extrapolation to $T=0$ seems to be reasonable and would
yield a critical chemical potential of $\mu_q(T=0)\approx 250~\mbox{MeV}$ 
or equivalently $\mu_B/T_c(\mu_B=0)\approx 4.7$.
This number appears to be at the lower edge of the phenomenological expectation
of $\mu_B/T_c(\mu_B=0)\approx 5-10$. Note, that our lattice spacing is 
close to the strong coupling regime and we should feel the influence of the
strong coupling limit. Strong coupling expansion calculations in general yield
much lower values of $\mu_B/T_c(\mu_B=0)\lsim 1.5$
\cite{Kawamoto:2005mq}.

For this work the density of state method has been employed, which works
well on small lattices up to chemical potentials of $\mu_q/T\lsim 3$ (other
methods \cite{Fodor:2004nz, Gavai:2003mf, methods} worked up to $\mu_q/T\lsim 1$). The
method is however extremely expensive and thus will in the near future not
yield results close to the thermodynamic limit or the continuum limit, due
to limitations in computer resources.

We have to emphasize that this simulations have been carried out on
coarse lattices with an unphysical value of $n_f=4$ degenerate fermion flavor,
and that neither the continuum nor the thermodynamic limit has been
taken. Since we used unimproved staggered fermions, the corrections due to a
finite lattice spacing are large. We also expect corrections due to the finite
size of our volume. The simulations have not been performed with
a constant quark mass, but $m_q/T=0.3$ has been held fixed.

\section*{Acknowledgments}
I would like to thank F. Karsch and Z. Fodor for helpful discussions and
comments. All members of the RBC-Bielefeld Collaboration are gratefully
acknowledged for providing me with preliminary data. This work has been supported
by the U.S. Department of Energy under contract DE-AC02-98CH1-886.

\end{document}